Review

# A Primer on Spacetime Singularities II: Spatially Homogeneous Cosmologies

**Jean-Pierre Luminet**

Laboratoire d'Astrophysique de Marseille (LAM), Centre National de La Recherche Scientifique, Aix-Marseille Université; jean-pierre.luminet@lam.fr

**Abstract**

This article is the second part of a review devoted to spacetime singularities in classical General Relativity. While Primer I developed the mathematical framework for defining, classifying, and proving the existence of singularities, the present work examines their concrete realization in spatially homogeneous cosmological models. After reviewing the geometrical foundations of homogeneity and the Bianchi classification, we derive the field equations for orthogonal and tilted perfect-fluid cosmologies and use the Kasner and Bianchi-I solutions as elementary prototypes of anisotropic singular behaviour. Raychaudhuri focusing provides general conditions under which the homogeneous Cauchy development reaches a finite-proper-time boundary. We then distinguish matter singularities, characterized by divergent density and Ricci curvature, from more subtle non-scalar or 'whimper' singularities associated with Cauchy horizons and extreme fluid tilt. The classical Ellis–King analysis is reconsidered in the light of later developments, including rigorous curvature-blow-up results for Bianchi VIII and IX models, kinematic singularities in future geodesically complete tilted spacetimes, and the recent reappearance of whimper behaviour in dipole cosmology. The resulting picture shows that even spatially homogeneous cosmologies exhibit a remarkably rich hierarchy of singular boundaries. The detailed dynamical approach to these boundaries—Hamiltonian cosmology, Kasner transitions, Mixmaster dynamics and the BKL conjecture—is reserved for Primer III.

## Introduction

The occurrence of singularities is one of the most robust predictions of classical General Relativity, but the singularity theorems themselves say remarkably little about the physical or geometrical nature of the singular boundary. Their essential conclusion is geodesic incompleteness: under appropriate global, causal and energy assumptions, some causal geodesics cannot be continued to arbitrary values of their affine parameter. Whether the corresponding boundary is characterized by divergent matter density, unbounded scalar curvature, a weaker non-scalar curvature pathology, a Cauchy horizon, or some other failure of regular evolution is a separate question.

The first article in this series (Luminet, 2025) was devoted to precisely this general background. It reviewed the definition of singularities through geodesic incompleteness, the difficulties involved in constructing singular boundaries, the distinction between quasi-regular, non-scalar and scalar curvature singularities, and the principal singularity theorems of Penrose and Hawking. The discussion was deliberately restricted to classical General Relativity and stopped short of analysing the internal structure and asymptotic dynamics of cosmological singularities.

The present article takes the next step by restricting attention to spatially homogeneous cosmologies. This specialization is strong enough to reduce Einstein's field equations essentially to systems of ordinary differential equations, yet weak enough to preserve a rich variety of relativistic phenomena absent from the isotropic Friedmann–Lemaître–Robertson–Walker (FLRW) models: anisotropic expansion, shear, nontrivial spatial curvature, fluid tilt and vorticity, Cauchy horizons, and several distinct kinds of singular boundary. Spatially homogeneous models therefore provide a particularly useful intermediate laboratory between the highly symmetric FLRW solutions and fully inhomogeneous cosmological spacetimes. Classical treatments were developed principally during the 1960s and 1970s and were synthesized in works such as Ellis and MacCallum (1969), Ryan and Shepley (1975), and Collins and Ellis (1979).

The geometrical organization of these models is supplied by the Bianchi classification of three-dimensional Lie algebras. A spacelike hypersurface on which a three-dimensional isometry group acts simply transitively may be assigned one of the Bianchi types, and the algebraic structure constants then determine the possible forms of its intrinsic spatial curvature. The original nineteenth-century classification by Bianchi, together with the later formulations adapted to relativistic cosmology by Taub, Behr, Ellis and MacCallum, remains the natural starting point for the subject.

A second fundamental distinction concerns the relation between the cosmological matter flow and the homogeneous hypersurfaces. In an orthogonal model the fluid four-velocity coincides with the unit normal to the hypersurfaces of homogeneity. In a tilted model it does not. This apparently modest generalization produces a remarkable enlargement of the possible behaviour: the fluid may have acceleration and vorticity, quantities homogeneous in the geometrically preferred frame acquire spatial gradients in the fluid rest frame, and the relative Lorentz factor between the two congruences may itself become singular. The systematic theory of tilted homogeneous cosmologies initiated by King and Ellis (1973) therefore occupies a central place in the discussion.

The first objective of this article is to determine under what circumstances spatially homogeneous cosmologies necessarily develop genuine curvature singularities. The Kasner and Bianchi-I solutions provide the elementary prototype, showing explicitly how anisotropic shear can dominate ordinary perfect-fluid matter near a singularity. More general results follow from the Raychaudhuri equation: under suitable energy conditions, the regular homogeneous Cauchy development must reach a focusing boundary after finite proper time. Additional use of the conservation laws and Einstein equations then often establishes that this boundary is a matter singularity, with divergent density and Ricci curvature. The exceptional stiff-fluid case already indicates, however, that the role of matter close to the singularity depends sensitively on its equation of state.

The second objective is to study situations in which focusing does not lead to divergent scalar curvature. The tilted Class-B cosmologies discovered in the classical literature provide striking examples. Ellis and King (1974) showed that the homogeneous Cauchy development may terminate at an intermediate or whimper singularity, associated with a Cauchy horizon while density and scalar curvature invariants remain finite. Collins (1974) subsequently demonstrated the strong dependence of such behaviour on the fluid equation of state, and the later review by Collins and Ellis (1979) established the whimper as a genuine alternative within Bianchi cosmology.

Far from having become merely an historical curiosity, this phenomenon has acquired renewed relevance. Coley, Hervik, Lim and MacCallum (2009) showed that extreme tilt may generate kinematic singularities at finite fluid proper time even when the underlying spacetime is future geodesically complete and its Ricci and Weyl curvature components remain bounded. More recently, Allahyari et al. (2025) found a whimper branch in tilted Bianchi-V/VIIh dipole cosmology, in which the tilt diverges while scalar curvature invariants remain finite. At the same time, rigorous mathematical results have considerably sharpened the opposite, generic curvature-blow-up picture for important homogeneous families, notably Bianchi VIII and IX (Ringström, 2000, 2001).

The article is organized as follows. Section 1 introduces the symmetry structure of spatially homogeneous spacetimes, invariant spatial frames and the Bianchi classification. Section 2 derives the field equations for orthogonal and tilted perfect-fluid cosmologies and examines the Kasner and Bianchi-I models. Section 3 turns to singularity formation itself: finite-time focusing of the homogeneous Cauchy development, matter singularities, intermediate and kinematic singularities, and finally the global Ellis–King classification of bang and whimper evolutions.

A deliberate boundary is placed at that point. The existence and geometrical nature of a singularity are logically distinct from the detailed dynamical process by which the spacetime approaches it. Kasner epochs and transitions, Hamiltonian cosmology, Bianchi VIII and IX oscillations, Mixmaster chaos, and the Belinskii–Khalatnikov–Lifshitz (BKL) conjecture constitute a subject of sufficient scope to require a separate treatment. These issues, together with the substantial developments that have occurred since the original work of the 1960s and 1970s, will be the subject of *A Primer on Spacetime Singularities III*.

## 1. Classification of Homogeneous Universes

### *1.1. Isometries of Spacetime: Basic Reminders*

In General Relativity, physical symmetries are naturally expressed as geometric symmetries of spacetime. A spacetime is said to be homogeneous with respect to a group of isometries if the group acts transitively on the relevant manifold or submanifold: any point can be mapped onto any other point by an isometry.

The strongest possible form of homogeneity is spacetime homogeneity, in which all events are geometrically equivalent. More relevant to cosmology is spatial homogeneity, where a three-dimensional group of isometries acts transitively on a family of spacelike hypersurfaces. All points belonging to a given hypersurface are then geometrically equivalent, although different hypersurfaces may correspond to different stages of cosmological evolution.

Isotropy concerns the equivalence of directions rather than points. At a point p, the subgroup of isometries leaving p fixed is called the isotropy group. If the isometry group acts transitively, the isotropy groups at different points are mutually isomorphic. In cosmology, local rotational symmetry refers to the existence of a preferred spatial direction about which all directions in the orthogonal two-plane are equivalent. Spherical symmetry means invariance under an SO(3) action whose generic orbits are two-spheres. At a generic point the isotropy subgroup preserves the radial direction and acts on the orthogonal two-plane; full equivalence of all spatial directions occurs only at a symmetry centre or in a genuinely isotropic homogeneous geometry.

Let $G_r$ denote an $r$-dimensional group of isometries acting on an $n$-dimensional manifold $M$. A group action is said to be *simply transitive* if, for every pair of points $p, q \in M$, there exists one and only one group transformation mapping $p$onto $q$. Otherwise, if the action is transitive but the transformation is not unique, it is multiply transitive.

An $n$ -dimensional (pseudo-)Riemannian manifold admits at most n(n+1)/2 independent Killing vectors. The maximum is attained precisely for spaces of constant sectional curvature. Thus a four-dimensional maximally symmetric spacetime possesses ten Killing vectors. Minkowski, de Sitter, and anti-de Sitter spacetime are the three standard Lorentzian examples, distinguished by the sign of the curvature.

A constant-curvature spacetime satisfies

$$R_{ab} = \lambda g_{ab},$$

where the curvature constant is spatially and temporally constant. The weaker Einstein-space condition R_ab = λ g_ab is not, by itself, sufficient for maximal symmetry in four dimensions, since the Weyl tensor may be non-zero. Maximal symmetry requires constant sectional curvature; equivalently, in four dimensions, an Einstein metric must also have vanishing Weyl tensor. In Einstein's equations this geometrical term can equivalently be interpreted as a cosmological constant or vacuum-energy contribution, for which

$$p = -\rho.$$

This case therefore has considerable physical significance in modern cosmology, particularly in connection with inflation and the observed accelerated expansion.

The Friedmann–Lemaître–Robertson–Walker (FLRW) models possess a six-dimensional group $G_6$ acting transitively on three-dimensional spacelike hypersurfaces of constant curvature. At every point of such a hypersurface there is a three-dimensional isotropy group. The six Killing vectors may therefore be understood as expressing three degrees of spatial homogeneity and three degrees of isotropy.

Spatially homogeneous but anisotropic cosmologies are less symmetric. Their homogeneous spacelike hypersurfaces admit a three-dimensional simply transitive group $G_3$ in the generic case. These models form the principal subject of this article.

There exists another class of spatially homogeneous models for which no three-dimensional subgroup acts simply transitively on the homogeneous hypersurfaces. The best-known examples are the Kantowski–Sachs models, whose symmetry group contains two-dimensional orbits of constant positive curvature. We shall return briefly to them in Section 3.2, but otherwise restrict the discussion to cosmologies admitting a simply transitive $G_3$ on their spatial hypersurfaces.

The classification of such $G_3$ groups is equivalent to the classification of three-dimensional real Lie algebras and leads to the Bianchi classification discussed in Section 1.6.

### *1.2. Completely Homogeneous Universes*

Spacetimes in which an isometry group acts transitively on the full four-dimensional manifold are *spacetime homogeneous*. Classical examples include the Einstein static universe and the Gödel universe. They are useful geometrical models, although their properties differ markedly from those required of an expanding cosmology.

Consider a spacetime-homogeneous model containing a perfect fluid,

$$T_{ab} = (\rho + p)u_a u_b + p\, g_{ab}. \qquad (1.1)$$

Homogeneity implies that the scalar thermodynamic variables are constant throughout spacetime,

$$\nabla_a \rho = 0, \nabla_a p = 0.$$

The conservation equations

$$\nabla_b T^{ab} = 0$$

may be decomposed into an energy equation and an Euler equation:

$$\dot{\rho} + (\rho + p)\theta = 0, \qquad (1.2a)$$

$$(\rho + p)\dot{u}_a + h^{ab}\nabla_b p = 0, \qquad (1.2b)$$

where $\theta = \nabla_a u^a$is the expansion scalar and $\dot{u}_a = u_{a;b}u^b$is the fluid acceleration.

For

$$\rho + p \neq 0,$$

spacetime homogeneity therefore implies

$$\theta = 0, \dot{u}_a = 0.$$

Thus the fluid does not undergo an overall expansion or contraction, and its worldlines are geodesic. The exceptional case $\rho + p = 0$corresponds to a vacuum-energy equation of state and must be treated separately.

The standard spacetime-homogeneous perfect-fluid solutions consequently do not exhibit the expanding Big-Bang-type behaviour that is the central concern of the present study. We shall therefore not consider this class further and turn instead to spacetimes that are homogeneous only on three-dimensional spatial sections.

### *1.3. The Friedmann–Lemaître–Robertson–Walker Models*

On sufficiently large scales, observations indicate a remarkable statistical isotropy of the Universe, most strikingly in the cosmic microwave background, but also in the large-scale distribution of matter. Combined with the Copernican assumption that our position is not privileged, this motivates the cosmological principle: at large scales, spatial hypersurfaces are taken to be both homogeneous and isotropic.

The corresponding spacetime possesses a six-dimensional group of spatial isometries, and the homogeneous hypersurfaces have constant curvature. In suitably chosen coordinates, the metric can be written

$$ds^2 = -dt^2 + R^2(t)d\Sigma^2, \qquad (1.3)$$

where $R(t)$is the cosmological scale factor and $d\Sigma^2$is the metric of a three-space of constant normalized curvature

$$K = +1, 0, -1.$$

A convenient form is

$$d\Sigma^2 = d\chi^2 + f_K^2(\chi)(d\theta^2 + \sin^2\theta\, d\phi^2), \qquad (1.4)$$

with

$$f_K(\chi) = \begin{cases} \sin\chi, & K = +1, \quad 0 \leq \chi \leq \pi, \\ \chi, & K = 0, \\ \sinh\chi, & K = -1. \end{cases} \qquad (1.5)$$

For the simply connected models, $K = +1$ corresponds to $S^3$, $K = 0$ to Euclidean space $E^3$, and $K = -1$ to hyperbolic space $H^3$.

It is important, however, not to confuse local curvature with global topology. The metric determines the local geometry but does not uniquely determine the global topology of the spatial sections. Constant-curvature spaces may be quotiented by discrete, fixed-point-free groups of isometries. Thus spatial sections of zero or negative curvature need not be infinite: compact flat and compact hyperbolic three-manifolds exist, while complete positive-curvature three-manifolds are compact spherical space forms. This distinction between geometry and topology has been discussed extensively in the cosmological context (see, for example, Lachièze-Rey and Luminet, 1995).

Spatial homogeneity implies that the density and pressure of a perfect cosmological fluid depend only on cosmic time:

$$\rho = \rho(t), p = p(t).$$

The fluid four-velocity is orthogonal to the homogeneous hypersurfaces, and therefore its acceleration, vorticity and shear vanish:

$$\dot{u}_a = 0, \omega_{ab} = 0, \sigma_{ab} = 0.$$

Its expansion scalar is

$$\theta = 3\frac{\dot{R}}{R}.$$

Unless otherwise specified, we shall set the cosmological constant to zero in the classical results discussed below. The conservation equation then becomes

$$\dot{\rho} = -3\,\frac{\dot{R}}{R}(\rho + p), \tag{1.6}$$

while the Raychaudhuri equation, or equivalently the acceleration Friedmann equation, gives

$$3\frac{\ddot{R}}{R} = -\frac{1}{2}(\rho + 3p). \tag{1.7}$$

Here and throughout the paper we use geometrized units $8\pi G = c = 1$.

If the strong energy condition is satisfied,

$$\rho + 3p \geq 0, \tag{1.8}$$

then

$$\ddot{R} \leq 0.$$

For an expanding model with $\dot{R} > 0$at some time $t_1$, extrapolation toward the past therefore leads to $R = 0$after a finite interval of proper time. More precisely, the concavity of $R(t)$ implies that the past endpoint occurs no farther away than the inverse instantaneous Hubble parameter,

$$t_1 - t_0 \leq H^{-1}(t_1), H = \frac{\dot{R}}{R}. \tag{1.9}$$

This conclusion depends essentially on the energy condition. It does **not** apply, for example, to a pure de Sitter phase, for which $p = -\rho$and the strong energy condition is violated.

To determine whether $R = 0$corresponds merely to a coordinate boundary or to a true curvature singularity, one must examine the matter and curvature invariants. Suppose, in addition to the strong energy condition, that the fluid satisfies the dominant-energy-type bound

$$-\frac{\rho}{3} \leq p \leq \rho. \tag{1.10}$$

Equation (1.6) then yields

$$-6 \leq \frac{d\ln\,\rho}{d\ln\,R} \leq -2,$$

and consequently, relative to any reference epoch $R_1$,

$$\rho_1\left(\frac{R_1}{R}\right)^2 \leq \rho(R) \leq \rho_1\left(\frac{R_1}{R}\right)^6. \tag{1.11}$$

Thus

$$\rho \to \infty \text{ as } R \to 0.$$

For a perfect fluid with vanishing cosmological constant,

$$R_{ab}R^{ab} = \rho^2 + 3p^2, \tag{1.12}$$

and this scalar curvature invariant therefore diverges. Since the Weyl tensor vanishes identically in FLRW spacetime, the singularity is entirely of Ricci type. Under the assumptions stated above, the past endpoint is consequently both geodesically incomplete and a scalar curvature singularity.

The important point is that this conclusion should not be confused with the observational status of the hot Big-Bang model. The cosmic microwave background, now observed as an almost perfect blackbody at approximately $2.7$ K, was released at recombination hundreds of thousands of years after the putative classical singular boundary. It provides compelling evidence for a hot, dense early Universe, but not direct observational evidence for the mathematical singularity obtained by extrapolating classical General Relativity to $R = 0$.

The existence of this singularity in the simplest homogeneous and isotropic cosmologies motivates the study of less symmetric models. There is no compelling reason to assume that the very early Universe was exactly isotropic, and anisotropic degrees of freedom may dominate or qualitatively modify the approach to the singular regime.

A natural first generalization is therefore to retain spatial homogeneity while abandoning isotropy. This leads to the Bianchi cosmologies and to the central questions addressed in the remainder of this article: do spatially homogeneous anisotropic universes generically possess singularities, what form do these singularities take, and how are they affected by the motion and physical properties of matter?

The detailed dynamical approach to such singularities—including Kasner transitions, Hamiltonian methods, Mixmaster dynamics and the BKL picture—will be reserved for the subsequent article in this series.

### *1.4. Choice of A Timelike Congruence*

By definition, a spatially homogeneous spacetime admits a foliation by spacelike hypersurfaces $S(t)$ on which a group of isometries $G_r$, with $r \geq 3$, acts transitively. Let $n^a$ denote the future-directed unit normal to these hypersurfaces,

$$n^a n_a = -1.$$

At least locally, the parameter $t$ may be chosen as proper time along the timelike geodesics orthogonal to one of the homogeneous hypersurfaces. Gaussian normal coordinates then give

$$n_a = -\nabla_a t, \tag{1.13}$$

and

$$n^b \nabla_b n^a = 0. \tag{1.14}$$

Thus the congruence defined by $n^a$ is geodesic and, being hypersurface-orthogonal, has vanishing vorticity:

$$\dot{n}^a = 0, \omega_{ab}[n] = 0. \tag{1.15}$$

Let $\{\xi_\alpha\}$ denote Killing vector fields generating the isometry group on the homogeneous hypersurfaces. They satisfy

$$\nabla_{(a}\xi_{\alpha\, b)} = 0 \tag{1.16}$$

and are tangent to $S(t)$:

$$n_a \xi_\alpha^a = 0. \tag{1.17}$$

The orthogonality is preserved along the normal geodesics. Indeed,

$$n^b \nabla_b (n_a \xi_\alpha^a) = \xi_\alpha^a n^b \nabla_b n_a + n^a n^b \nabla_b \xi_{\alpha a} = 0, \tag{1.18}$$

where the first term vanishes because the normal congruence is geodesic and the second because of the Killing equation. Hence a hypersurface obtained by translating $S(t)$a fixed proper time along the normal congruence is again orthogonal to $n^a$.

The time coordinate may consequently be chosen so that the homogeneous surfaces are simply

$$S(t): t = \text{const},$$

and in synchronous coordinates

$$n^a = \left(\frac{\partial}{\partial t}\right)^a. \tag{1.19}$$

Since the isometries map homogeneous hypersurfaces into themselves and preserve their unit normal, $n^a$ is invariant under the spatial symmetry group:

$$\mathcal{L}_{\xi_\alpha} n^a = [\xi_\alpha, n]^a = 0. \tag{1.20}$$

This geometrically preferred timelike congruence must be distinguished from the congruence defined by matter. If the universe contains a fluid with four-velocity $u^a$, spatial homogeneity requires $u^a$also to be invariant under the symmetry group,

$$\mathcal{L}_{\xi_\alpha} u^a = 0. \tag{1.21}$$

Two distinct situations can then occur. If

$$u^a = n^a, \tag{1.22}$$

the fluid worldlines are orthogonal to the homogeneous hypersurfaces and the cosmological model is called orthogonal. If

$$u^a \neq n^a, \quad (1.23)$$

the fluid has a nonzero velocity relative to the homogeneous hypersurfaces and the model is called *tilted*.

This distinction is physically important. The normal congruence $n^a$ is always vorticity-free and geodesic in the synchronous representation adopted here, whereas a tilted fluid congruence $u^a$ may possess acceleration and vorticity. We shall return to this situation in Section 2.2.

### *1.5. Invariant Spatial Frames*

The geometry of a spatially homogeneous spacetime is most conveniently described using a spatial frame adapted to the symmetry group.

Let $G_3$ act simply transitively on each homogeneous hypersurface, and let its Killing generators $\xi_\alpha$ satisfy

$$[\xi_\alpha, \xi_\beta] = C^\gamma_{\alpha\beta}\xi_\gamma, \quad (1.24)$$

where the $C^\gamma_{\alpha\beta}$ are the structure constants of the corresponding Lie algebra.

Because the group action is simply transitive, one may construct another spatial frame $\{X_\alpha\}$, tangent to the homogeneous hypersurfaces and invariant under the action generated by the Killing fields:

$$[X_\alpha, \xi_\beta] = 0. \quad (1.25)$$

With the convention adopted here, the commutators of this invariant frame are

$$[X_\alpha, X_\beta] = -C^\gamma_{\alpha\beta}X_\gamma. \quad (1.26)$$

The minus sign reflects the familiar relation between left- and right-invariant vector fields on a Lie group.

The invariant frame may be propagated along the normal congruence so that

$$[n, X_\alpha] = 0. \quad (1.27)$$

In the synchronous coordinates introduced above, this simply means that the components of the $X_\alpha$ can be chosen independent of $t$.

Spatial homogeneity then implies that the scalar products

$$g_{\alpha\beta}(t) \equiv g_{ab}X^a_\alpha X^b_\beta \quad (1.28)$$

are constant on every homogeneous hypersurface and therefore depend on time only.

Let $\{\omega^\alpha\}$ be the dual basis of invariant one-forms,

$$\omega^\alpha(X_\beta) = \delta^\alpha_\beta. \quad (1.29)$$

With the convention of Eq. (1.26), they satisfy the Maurer–Cartan relations

$$d\omega^\alpha = \frac{1}{2}C^\alpha_{\beta\gamma}\,\omega^\beta \wedge \omega^\gamma. \quad (1.30)$$

The general spatially homogeneous metric can therefore be written locally in the compact form

$$\boxed{ds^2 = -dt^2 + g_{\alpha\beta}(t)\,\omega^\alpha\omega^\beta} \quad (1.31)$$

when the normal congruence is used as the time direction.

This expression separates very clearly the two ingredients entering a Bianchi cosmology. The invariant one-forms $\omega^\alpha$, through their structure constants $C^\alpha_{\beta\gamma}$, determine the algebraic type of spatial homogeneity, whereas the six functions $g_{\alpha\beta}(t)$ describe its dynamical evolution.

Alternative frame choices are possible. One may use an orthonormal triad together with $n^a$, which is particularly convenient for writing Einstein's equations in first-order form. In a fluid spacetime one may instead take $u^a$ as the timelike basis vector and construct either an invariant or an orthonormal spatial frame relative to it. In tilted models, however, the instantaneous rest spaces orthogonal to $u^a$ do not coincide with the homogeneous hypersurfaces.

For the purposes of the present article, the two most useful descriptions will therefore be:

$$\{n^a, X^a_\alpha\},$$

which makes spatial homogeneity explicit, and an orthonormal tetrad adapted either to $n^a$ or, when appropriate, to the fluid velocity $u^a$.

The classification of spatially homogeneous cosmologies now reduces to the algebraic classification of the possible structure constants $C^\alpha_{\beta\gamma}$.

*1.6. Bianchi Classification of Three-Dimensional Homogeneity Groups*

The simply transitive three-dimensional isometry groups relevant to spatially homogeneous cosmology are classified by their real three-dimensional Lie algebras. The original classification is due to Bianchi (1898), with forms particularly adapted to cosmology developed by Taub, Behr, Ellis and MacCallum (see Estabrook, Wahlquist and Behr, 1968; Ellis and MacCallum, 1969).

Let

$$[\xi_\alpha, \xi_\beta] = C^\gamma_{\alpha\beta}\, \xi_\gamma. \quad (1.32)$$

The structure constants satisfy

$$C^\gamma_{\alpha\beta} = -C^\gamma_{\beta\alpha}. \quad (1.33)$$

and the Jacobi identities

$$C^\epsilon_{\alpha\beta} C^\delta_{\gamma\epsilon} + C^\epsilon_{\beta\gamma} C^\delta_{\alpha\epsilon} + C^\epsilon_{\gamma\alpha} C^\delta_{\beta\epsilon} = 0. \quad (1.34)$$

In three dimensions, the structure constants can be decomposed uniquely into a symmetric tensor $n^{\alpha\beta}$and a vector $a_\alpha$:

$$C^\gamma{}_{\alpha\beta} = \epsilon_{\alpha\beta\delta} n^{\delta\gamma} + a_\alpha \delta^\gamma{}_\beta - a_\beta \delta^\gamma{}_\alpha \quad (1.35)$$

with

$$n^{\alpha\beta} = n^{\beta\alpha}, a_\alpha = \tfrac{1}{2} C^\beta_{\alpha\beta}. \quad (1.36)$$

With this convention, contraction of Eq. (1.35) gives $C^\beta{}_{\alpha\beta}=2a_\alpha$, consistently with Eq. (1.36). This convention will be used throughout the paper.

The Jacobi identities reduce to the remarkably simple algebraic condition

$$n^{\alpha\beta} a_\beta = 0. \quad (1.37)$$

This immediately leads to the fundamental division of Bianchi models into two classes:

$$\text{Class A}: a_\alpha = 0, \quad (1.38a)$$

$$\text{Class B}: a_\alpha \neq 0. \quad (1.38b)$$

In Lie-group terminology, Class A consists of the unimodular three-dimensional Lie groups, while Class B consists of the non-unimodular ones.

Because $n^{\alpha\beta}$ is real and symmetric, a linear transformation of the invariant frame can be used to diagonalize it. For Class A we may therefore choose

$$n^{\alpha\beta} = \text{diag}\,(n_1, n_2, n_3), \quad (1.39)$$

and rescale the frame so that each nonzero $n_i$ is +1or −1. This gives the six Class-A Bianchi types:

| **Bianchi type** | $(n_1, n_2, n_3)$ |
|---|---|
| I | (0, 0, 0) |
| II | (1, 0, 0) |
| $VI_0$ | $(1, -1, 0)$ |
| $VII_0$ | $(1,1,0)$ |

| **Bianchi type** | $(n_1, n_2, n_3)$ |
|---|---|
| VIII | $(1,1,-1)$ |
| IX | $(1,1,1)$ |

**Table 1**

For Class B, Eq. (1.36) implies that $a_\alpha$ lies in the null space of $n^{\alpha\beta}$. Retaining the convention used throughout the present article, we choose

$$a_\alpha = (0,0,a), \quad n^{\alpha\beta} = \mathrm{diag}\,(n_1, n_2, 0). \tag{1.40}$$

The canonical forms may then be chosen as follows:

| **Bianchi type** | $(n_1, n_2, n_3)$ | $a$ |
|---|---|---|
| V | $(0,0,0)$ | 1 |
| IV | $(1,0,0)$ | 1 |
| VI$_h$, $h < 0$ | $(1,-1,0)$ | $\sqrt{-h}$ |
| VII$_h$, $h > 0$ | $(1,1,0)$ | $\sqrt{h}$ |

**Table 2**

The continuous parameter $h$ is invariant under admissible changes of basis. In the present convention it is defined by

$$h = \frac{a^2}{n_1 n_2} \tag{1.41}$$

when $n_1 n_2 \neq 0$. Thus $h < 0$ for type VI$_h$ and $h > 0$ for type VII$_h$. Bianchi type III corresponds to the special case

$$\mathrm{III} = \mathrm{VI}_{-1}. \tag{1.42}$$

The literature contains several slightly different conventions for the parameter $h$, arising from changes in signs, ordering of basis vectors or normalization of the structure constants. Equation (1.41) fixes the convention used here. This point will become particularly relevant later for the exceptional model $\mathrm{VI}_{-1/9}$.

The complete Bianchi classification may therefore be summarized as

$$\boxed{\begin{array}{ll}\text{Class A:} & \mathrm{I, II, VI_0, VII_0, VIII, IX,}\\ \text{Class B:} & \mathrm{V, IV, VI}_h, \mathrm{VII}_h.\end{array}} \tag{1.43}$$

Some of these algebras have particularly familiar group-theoretic interpretations. Type I is the Abelian translation group $\mathbb{R}^3$; Type II is the three-dimensional Heisenberg group; Type VIII is locally associated with $SL(2,\mathbb{R})$; and Type IX has Lie algebra

$$\mathfrak{su}(2) \simeq \mathfrak{so}(3).$$

The simply connected Bianchi-IX group itself is

$$SU(2) \simeq S^3,$$

rather than $SO(3)$. This distinction between a Lie algebra and a particular global Lie group realization is important. More generally, the Bianchi classification is fundamentally local and algebraic; spatial sections may possess nontrivial global topology obtained by taking appropriate discrete quotients of the simply connected covering spaces.

The relation with the FLRW models is especially instructive. The spatially flat FLRW geometry admits simply transitive subgroups of Bianchi types I and VII$_0$; the negatively curved FLRW geometry admits types V and VII$_h$; and the positively curved FLRW geometry is associated with type IX. Thus the isotropic cosmologies appear as special highly symmetric members of particular Bianchi families.

The Bianchi types are not completely isolated from one another in the space of Lie-algebra structures. Certain types occur as limiting or boundary cases of more general ones; these specialization relations are represented in Figure 1. They will become useful later when discussing the dynamical behavior of homogeneous cosmologies and, in the subsequent article of this series, the succession of asymptotic Bianchi states near a cosmological singularity.

Explicit coordinate realizations of the Killing vectors, invariant vector bases and dual one-forms can be constructed for every Bianchi type. These formulae are standard and will not be required in the subsequent analysis; we therefore do not reproduce them here.

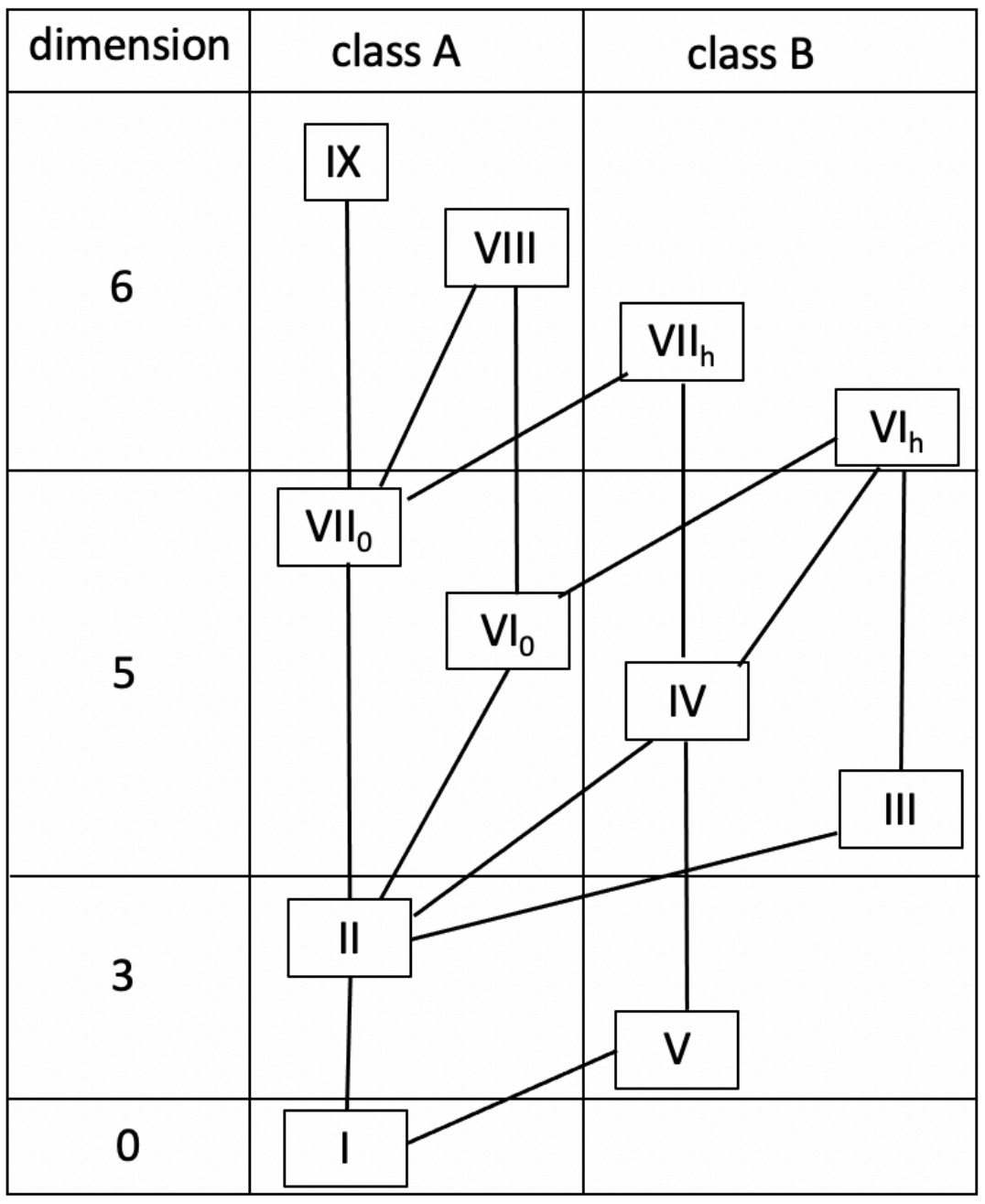

**Figure 1**: Specialization diagram; a line descending from one type to another indicates that the second type is contained within the boundary of the first.

## 2. Spatially Homogeneous Cosmologies

*2.1. Field Equations for Spatially Homogeneous Models*

We now apply the geometrical framework of Section 1 to the Einstein equations. In the synchronous frame defined by the unit normals $n^a$ to the homogeneous hypersurfaces, the general spatially homogeneous metric was written as

$$ds^2 = -dt^2 + g_{\alpha\beta}(t)\, \omega^\alpha \omega^\beta. \tag{2.1}$$

where the invariant one-forms satisfy

$$d\omega^\alpha = \frac{1}{2} C^\alpha_{\beta\gamma}\, \omega^\beta \wedge \omega^\gamma. \tag{2.2}$$

The six independent components of $g_{\alpha\beta}(t)$ describe the time evolution of the intrinsic geometry of the homogeneous hypersurfaces, while the constants $C^\alpha_{\beta\gamma}$ specify their Bianchi type.

***Expansion and anisotropy***

It is convenient to separate the overall expansion from the anisotropic degrees of freedom. Following Misner, we write

$$g_{\alpha\beta} = e^{2\alpha(t)} \left(e^{2\beta(t)}\right)_{\alpha\beta}, \qquad \mathrm{tr}\,\beta = 0. \tag{2.3}$$

The quantity

$$R(t) = e^{\alpha(t)} \tag{2.4}$$

plays the role of an average scale factor, since the volume element of a homogeneous hypersurface is proportional to

$$\sqrt{g} = e^{3\alpha}. \tag{2.5}$$

The expansion tensor of the normal congruence is

$$\Theta_{ab} = h_a^c h_b^d \nabla_c n_d, \tag{2.6}$$

where

$$h_{ab} = g_{ab} + n_a n_b$$

is the induced spatial metric. It decomposes as

$$\Theta_{\alpha\beta} = \frac{1}{3}\theta\, h_{\alpha\beta} + \sigma_{\alpha\beta}, \tag{2.7}$$

where

$$\theta = \nabla_a n^a = 3\dot{\alpha} = 3\frac{\dot{R}}{R} \tag{2.8}$$

is the expansion scalar and $\sigma_{\alpha\beta}$ is the shear tensor,

$$\sigma_\alpha^\alpha = 0.$$

We use the standard abbreviation

$$\sigma^2 \equiv \frac{1}{2}\sigma_{\alpha\beta}\sigma^{\alpha\beta}. \tag{2.9}$$

If $\beta$ commutes with $\dot{\beta}$, one may choose a frame in which both are diagonal, and then

$$\sigma_\beta^\alpha = \dot{\beta}_\beta^\alpha. \tag{2.10}$$

In the general non-diagonal case, $\beta$ and $\dot{\beta}$ need not commute. A time-dependent orthonormal spatial frame then acquires an angular velocity relative to a Fermi-propagated frame. This distinction becomes important in the treatment of non-diagonal Bianchi models, but will not be needed explicitly in the present section.

Since the normal congruence is geodesic and hypersurface-orthogonal,

$$\dot{n}_a = 0, \qquad \omega_{ab}[n] = 0. \tag{2.11}$$

Thus its kinematics is completely characterized by $\theta$ and $\sigma_{\alpha\beta}$.

***Einstein equations for an orthogonal perfect fluid***

Consider first an orthogonal perfect fluid,

$$u^a = n^a,$$

with stress-energy tensor

$$T_{ab} = (\rho + p) n_a n_b + p\, g_{ab}. \tag{2.12}$$

As in Section 1.3, we set the cosmological constant to zero unless explicitly stated otherwise and use units $8\pi G = c = 1$.

The Einstein equations divide naturally into constraint equations intrinsic to each homogeneous hypersurface and evolution equations.

The Hamiltonian constraint is

$$R^{(3)} + \theta^2 - \Theta_{\alpha\beta}\Theta^{\alpha\beta} = 2\rho, \tag{2.13}$$

or, using Eq. (2.7),

$$\frac{1}{3}\theta^2 = \sigma^2 + \rho - \frac{1}{2}R^{(3)}, \tag{2.14}$$

where $R^{(3)}$ is the scalar curvature of the homogeneous spatial hypersurfaces.

Equation (2.14) generalizes the first Friedmann equation. Besides the matter density, the expansion is now affected both by spatial curvature and by anisotropy.

The momentum constraints are

$$D_\beta\left(\Theta_\alpha^\beta - \delta_\alpha^\beta\theta\right) = 0. \tag{2.15}$$

Since all scalar quantities are spatially homogeneous,

$$D_\alpha\theta = 0,$$

so that

$$D_\beta\sigma_\alpha^\beta = 0. \tag{2.16}$$

In an orthonormal spatial frame $\{e_\alpha(t)\}$ adapted to the homogeneous hypersurfaces, define the time-dependent commutator coefficients $\gamma^\gamma{}_{\alpha\beta}(t)$ by $[e_\alpha(t), e_\beta(t)] = \gamma^\gamma{}_{\alpha\beta}(t) e_\gamma(t)$. These coefficients are generally time-dependent and must be distinguished from the fixed structure constants $C^{\gamma}_{\alpha\beta}$ of the Killing basis introduced in Eq. (1.32). At each time, their decomposition into $n^{\alpha\beta}(t)$ and $a_{\alpha}(t)$ uses the same sign convention as Eq. (1.35).

The momentum constraint then becomes the purely algebraic relation

$$3a^{\beta}\sigma_{\alpha\beta} + \epsilon_{\alpha\beta\gamma}n^{\beta}_{\delta}\sigma^{\gamma\delta} = 0. \qquad (2.17)$$

This equation places important restrictions on the relative orientation of the shear and the preferred directions defined by the Bianchi algebra.

The trace of the spatial Einstein equations yields the Raychaudhuri equation,

$$\dot{\theta} = -\tfrac{1}{3}\theta^2 - 2\sigma^2 - \tfrac{1}{2}(\rho + 3p), \qquad (2.18)$$

for the geodesic normal congruence.

The trace-free part gives the evolution of the shear. In a Fermi-propagated orthonormal frame it takes the particularly simple form

$$\dot{\sigma}_{\alpha\beta} + \theta\sigma_{\alpha\beta} + S^{(3)}_{\alpha\beta} = 0, \qquad (2.19)$$

where

$$S^{(3)}_{\alpha\beta} = R^{(3)}_{\alpha\beta} - \tfrac{1}{3}R^{(3)}\,\delta_{\alpha\beta} \qquad (2.20)$$

is the trace-free part of the Ricci tensor of the homogeneous three-space.

Equation (2.19) clearly exhibits the dynamical origin of anisotropy. In Bianchi I,

$$R^{(3)}_{\alpha\beta} = 0,$$

and the shear simply decays according to

$$\dot{\sigma}_{\alpha\beta} + \theta\sigma_{\alpha\beta} = 0.$$

For the other Bianchi types, anisotropic spatial curvature acts as a source for the shear. This interaction between shear and three-curvature will become central to the Hamiltonian and Mixmaster analysis developed in the subsequent article of this series.

Finally, conservation of stress-energy gives

$$\dot{\rho} + (\rho + p)\theta = 0. \qquad (2.21)$$

For a barotropic equation of state

$$p = (\gamma - 1)\rho, \qquad (2.22)$$

this integrates to

$$\rho = \rho_0 e^{-3\gamma\alpha} = \rho_0 R^{-3\gamma}. \qquad (2.23)$$

This relation will be particularly useful when comparing the asymptotic importance of matter and shear near a cosmological singularity.

***Spatial curvature and Bianchi type***

The intrinsic curvature of the homogeneous hypersurfaces is determined algebraically by the commutator variables $a_{\alpha}$ and $n_{\alpha\beta}$. Its scalar curvature is

$$R^{(3)} = -6a_{\alpha}a^{\alpha} - n_{\alpha\beta}n^{\alpha\beta} + \tfrac{1}{2}(n^{\alpha}_{\alpha})^2. \qquad (2.24)$$

No spatial derivatives occur because all geometrical quantities are homogeneous.

For Class A,

$$a_{\alpha} = 0,$$

whereas Class B possesses

$$a_{\alpha} \neq 0.$$

Bianchi I has

$$a_{\alpha} = 0, \qquad n_{\alpha\beta} = 0,$$

and therefore

$$R^{(3)} = 0 \qquad (2.25)$$

identically.

Type $\text{VII}_0$ also admits a flat three-geometry for appropriate choices of the spatial metric; indeed Euclidean three-space possesses simply transitive isometry subgroups of both types I and $\text{VII}_0$. Conversely, negatively curved FLRW space admits Bianchi V and $\text{VII}_h$ simply transitive subgroups, while the positively curved FLRW geometry corresponds to Bianchi IX.

One should not, however, infer the sign of $R^{(3)}$ from the Bianchi label alone in a general anisotropic metric. In particular, although the isotropic Bianchi-IX geometry has positive curvature, sufficiently anisotropic type-IX metrics need not have positive scalar curvature. What the Bianchi type fixes is the underlying Lie-algebra structure, not the instantaneous value or even, in all cases, the sign of the spatial scalar curvature.

***Diagonal and non-diagonal models***

The momentum constraints and the freedom to perform automorphisms of the Bianchi group considerably reduce the number of independent metric variables.

For the orthogonal perfect-fluid models considered here, the classical analysis of Ellis and MacCallum (1969) shows that Class-A models may be represented, after a suitable choice of invariant frame, by a diagonal spatial metric,

$$ds^2 = -dt^2 + A^2(t)(\omega^1)^2 + B^2(t)(\omega^2)^2 + C^2(t)(\omega^3)^2. \quad (2.26)$$

An analogous diagonal representation applies to the relevant Class-B orthogonal models satisfying the corresponding algebraic restrictions.

There is, however, an important exceptional Class-B case: Bianchi $\mathrm{VI}_{-1/9}$, for which the momentum constraints allow an additional non-diagonal degree of freedom. This exceptional model repeatedly appears in the classical classification of homogeneous cosmologies and has also acquired renewed interest in recent studies of their singular asymptotics. Its detailed dynamical significance will be discussed in *Primer III*.

The diagonal representation should not be confused with a statement that all homogeneous cosmologies are intrinsically diagonal. Once the fluid is tilted, or more general matter sources are introduced, genuinely non-diagonal solutions occur and may involve rotations of the preferred spatial axes.

This brings us naturally to the second major class of homogeneous cosmologies, in which the fluid velocity is not orthogonal to the hypersurfaces of homogeneity.

## *2.2. Spatially Homogeneous Tilted Models*

In the models considered in the preceding section, the fluid four-velocity $u^a$ was identified with the unit normal $n^a$ to the homogeneous hypersurfaces. This need not be the case. A spatially homogeneous geometry may contain matter whose worldlines are inclined with respect to these hypersurfaces. Such models are called 'tilted' spatially homogeneous cosmologies (King and Ellis, 1973). The original formulation in terms of the relative hyperbolic angle between $u^a$ and $n^a$ remains particularly useful.

Let

$$h_{ab} = g_{ab} + n_a n_b$$

denote the projection tensor onto the homogeneous hypersurfaces, and let

$$\tilde{h}_{ab} = g_{ab} + u_a u_b \quad (2.27)$$

project onto the instantaneous rest space of the fluid.

The fluid velocity can be decomposed relative to the normal congruence as

$$u^a = \Gamma(n^a + v^a), \qquad n_a v^a = 0, \qquad v^2 \equiv v_a v^a < 1, \quad (2.28)$$

where

$$\Gamma = (1 - v^2)^{-1/2}$$

is the Lorentz factor. Writing

$$v^a = v\, e^a, \qquad e_a e^a = 1, \qquad e_a n^a = 0,$$

we introduce the tilt rapidity $\beta$ by

$$v = \tanh\beta, \qquad \Gamma = \cosh\beta, \qquad \Gamma v = \sinh\beta. \quad (2.29)$$

Thus

$$u^a = \cosh\beta\, n^a + \sinh\beta\, e^a. \quad (2.30)$$

The non-tilted case corresponds to

$$\beta = 0, \qquad u^a = n^a.$$

For $\beta \neq 0$, the instantaneous rest spaces of the fluid do not coincide with the homogeneous hypersurfaces.

It is sometimes useful to introduce the unit spacelike vector $c^a$ lying in the plane spanned by $u^a$ and $n^a$, but orthogonal to the fluid velocity:

$$c^a = \sinh\beta\, n^a + \cosh\beta\, e^a, \qquad c_a u^a = 0, \qquad c_a c^a = 1. \tag{2.31}$$

The inverse relations are

$$n^a = \cosh\beta\, u^a - \sinh\beta\, c^a,$$
$$e^a = -\sinh\beta\, u^a + \cosh\beta\, c^a. \tag{2.32}$$

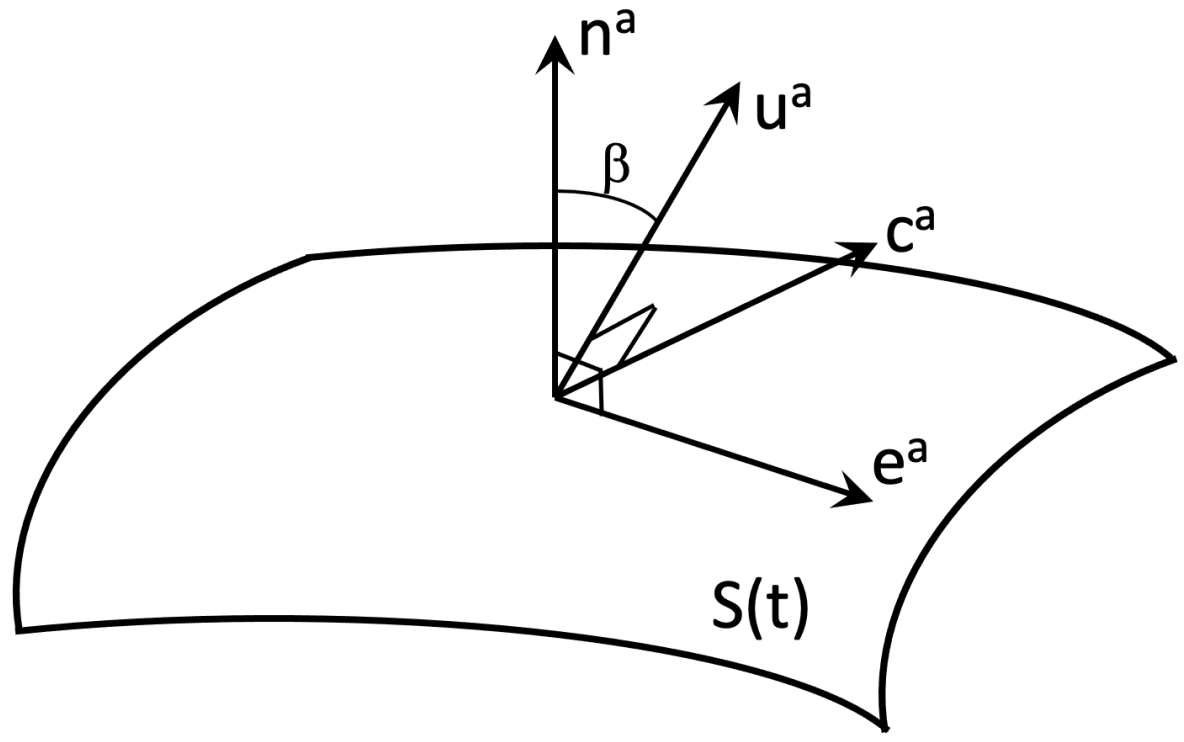

**Figure 2**: Geometry of the tilt decomposition. The fluid four-velocity is inclined by the rapidity β relative to the unit normal to the homogeneous hypersurface; the labelled spatial vectors are the associated unit directions.

Although the spacetime is homogeneous on the hypersurfaces $S(t)$, a tilted observer does not in general interpret the matter variables as spatially homogeneous in his or her instantaneous rest space. Indeed, for any scalar $f = f(t)$,

$$u^a \nabla_a f = \cosh\beta \, \frac{df}{dt},$$

whereas

$$\tilde{h}_a^b \nabla_b f = \sinh\beta \, \frac{df}{dt}\, c_a. \tag{2.33}$$

Thus a quantity such as $\rho(t)$, although constant over each homogeneous hypersurface, possesses a spatial gradient in the rest frame of a tilted fluid observer whenever $\beta \neq 0$ and $\dot{\rho} \neq 0$. This elementary fact is one of the characteristic features of tilted cosmology.

***Fluid conservation and acceleration***

For a perfect fluid,

$$T_{ab} = (\rho + p) u_a u_b + p\, g_{ab}. \tag{2.34}$$

The conservation equations

$$\nabla_b T^{ab} = 0$$

split, relative to $u^a$, into

$$u^a \nabla_a \rho + (\rho + p)\Theta = 0, \tag{2.35}$$

and

$$(\rho + p)\dot{u}_a + \tilde{h}_a^b \nabla_b p = 0, \tag{2.36}$$

where

$$\Theta = \nabla_a u^a$$

is the fluid expansion and

$$\dot{u}_a = u^b \nabla_b u_a$$

its acceleration.

For a barotropic equation of state $p = p(\rho)$, it is convenient, following the classical tilted-model analysis, to introduce two functions $r(\rho)$ and $s(\rho)$ by

$$\frac{dr}{r} = \frac{dp}{\rho+p}, \qquad \frac{ds}{s} = \frac{d\rho}{\rho+p}. \tag{2.37}$$

Their normalization may be chosen so that

$$rs = \rho + p. \tag{2.38}$$

Equation (2.35) then gives

$$\Theta = -\cosh\beta \, \frac{d\ln s}{dt}, \tag{2.39}$$

while the Euler equation becomes

$$\dot{u}_a = -\sinh\beta \, \frac{d\ln r}{dt} \, c_a. \tag{2.40}$$

Thus the acceleration, when nonzero, is directed along the tilt direction in the fluid rest frame. In particular, a tilted dust fluid ($p = 0$) is geodesic, whereas for a genuinely evolving barotropic fluid with $dp/d\rho \neq 0$, nonzero tilt generally produces nonzero acceleration.

This corrects a potentially misleading point in some early formulations of the tilted-fluid equations: pressure gradients that are absent on the homogeneous hypersurfaces are nevertheless present in the instantaneous rest spaces of tilted fluid observers.

The subset $\beta = 0$ is dynamically invariant. Consequently, under regular evolution a solution that is initially tilted does not cross into the exactly non-tilted subset at a finite time, although the tilt may of course tend asymptotically to zero. Conversely, an initially orthogonal perfect-fluid Bianchi model remains orthogonal.

***Stress-energy tensor in the normal frame***

Although the matter is a perfect fluid in its own rest frame, it does not appear as a perfect fluid to observers following the normal congruence $n^a$. Relative to $n^a$, write

$$T_{ab} = \mu n_a n_b + 2q_{(a}n_{b)} + Ph_{ab} + \pi_{ab}, \tag{2.41}$$

where $q_a n^a = 0$, $\pi_{ab} n^b = 0$, and $\pi^a_{\ a} = 0$.

Substitution of Eq. (2.30) gives

$$\mu = \rho + (\rho + p)\sinh^2\beta, \tag{2.42a}$$

$$q_a = (\rho + p)\sinh\beta\cosh\beta \, e_a, \tag{2.42b}$$

$$P = p + \frac{1}{3}(\rho + p)\sinh^2\beta, \tag{2.42c}$$

and

$$\pi_{ab} = (\rho + p)\sinh^2\beta \left(e_a e_b - \frac{1}{3}h_{ab}\right). \tag{2.42d}$$

Thus tilt generates, in the normal frame, an effective energy flux and anisotropic pressure even though the fluid itself is perfect.

The geometrical equations derived in Section 2.1 therefore remain applicable if the perfect-fluid variables $\rho$ and $p$ are replaced by the corresponding normal-frame quantities $\mu$, $P$, $q_a$ and $\pi_{ab}$.

The Hamiltonian constraint becomes

$$\frac{1}{3}\theta^2 = \sigma^2 + \mu - \frac{1}{2} R^{(3)}, \tag{2.43}$$

while the momentum constraint acquires the energy-flux source

$$D_b \sigma^b_{\ a} = -q_a. \tag{2.44}$$

The Raychaudhuri equation for the normal congruence is

$$\dot{\theta} = -\frac{1}{3}\theta^2 - 2\sigma^2 - \frac{1}{2}(\rho + 3p) - (\rho + p)\sinh^2\beta. \tag{2.45}$$

The additional non-negative term proportional to $\sinh^2\beta$, when $\rho + p \geq 0$, strengthens the focusing of the normal congruence.

Finally, the shear propagation equation becomes

$$\dot{\sigma}_{\alpha\beta} + \theta\sigma_{\alpha\beta} + S^{(3)}_{\alpha\beta} = \pi_{\alpha\beta}. \tag{2.46}$$

Tilt therefore influences the geometry not only through its contribution to the effective energy density but also through an effective anisotropic stress.

***Restrictions imposed by the Bianchi type***

The momentum constraint and the algebraic structure of the homogeneity group impose strong restrictions on the possible tilted solutions. King and Ellis derived a number of particularly useful results for perfect-fluid Bianchi cosmologies.

For Bianchi type I all spatial structure constants vanish. Equation (2.44) then implies $q_a = 0$; provided $\rho + p > 0$, Eq. (2.42b) gives

$$\beta = 0.$$

Thus there are no genuinely tilted single perfect-fluid Bianchi-I models under these assumptions.

Tilted type-II perfect-fluid models are vorticity-free, whereas genuinely tilted models of types VIII and IX necessarily possess fluid vorticity. For Class-B models the relation between the tilt direction and the preferred vector $a_\alpha$ introduced in Section 1.6 becomes important. If the tilt direction is parallel to $a_\alpha$, the fluid vorticity vanishes and this direction is generically an eigenvector of the shear of the normal congruence.

The exceptional case is again Bianchi $VI_{-1/9}$, for which an additional degree of freedom is permitted. If the tilt and $a_\alpha$ are not parallel, vorticity is generically nonzero, apart from special type-III solutions.

Another classical result is that a genuinely tilted perfect-fluid Bianchi model of the class considered by King and Ellis cannot be shear-free. The shear-free expanding homogeneous perfect-fluid limit corresponds to the isotropic FLRW case. These results illustrate the close relation between tilt, shear, vorticity and the algebraic structure of the Bianchi group.

***Modern perspective on tilt***

Tilt subsequently became an important part of the dynamical-systems treatment of Bianchi cosmology. A particularly significant possibility is 'extreme tilt', $v \to 1$ or equivalently $|\beta| \to \infty$, in which the fluid worldlines become asymptotically null relative to the homogeneous foliation. This can lead to a striking distinction between the geometrical behaviour of the spacetime and the experience of observers comoving with the fluid.

In particular, tilted LRS Bianchi-V models provide examples of spacetimes that are future geodesically complete while the fluid congruence reaches a finite-proper-time kinematic singularity; in suitable examples the components of the Ricci and Weyl tensors, and even arbitrarily many of their covariant derivatives, remain bounded (Coley et al., 2009). This development gives a modern interpretation to phenomena already encountered in the Ellis–King analysis.

More recently, tilted Bianchi $V/VII_h$ models have again attracted attention in "dipole cosmology." One lesson is that metric shear and fluid tilt are dynamically distinct: the shear may become small while the tilt grows, even in accelerating cosmologies. Even more directly relevant to the present article, recent analyses of the approach to the Big Bang have recovered an Ellis–King whimper branch in which the tilt diverges while the matter density and scalar curvature invariants can remain finite (Allahyari et al., 2025).

We shall return to these kinematic and whimper singularities in Section 3.3. For the moment, the important point is that tilt introduces a physically independent degree of freedom whose behaviour may differ radically from that of the metric anisotropy itself.

### *2.3. Kasner and Bianchi I Models*

The simplest anisotropic spatially homogeneous cosmologies are the Bianchi-I models, for which

$$C^{\gamma}_{\alpha\beta} = 0, \qquad a_\alpha = 0, \qquad n_{\alpha\beta} = 0. \tag{2.47}$$

Hence the homogeneous hypersurfaces are intrinsically flat,

$$R^{(3)}_{\alpha\beta} = 0, \qquad R^{(3)} = 0. \tag{2.48}$$

Bianchi I may therefore be regarded as the anisotropic generalization of the spatially flat FLRW universe. It provides the elementary prototype for much of the singular behaviour encountered later in more complicated homogeneous cosmologies.

In an invariant frame the metric may be written in diagonal form as

$$ds^2 = -dt^2 + A^2(t)\,dx^2 + B^2(t)\,dy^2 + C^2(t)\,dz^2. \tag{2.49}$$

It is convenient to introduce the average scale factor

$$R = (ABC)^{1/3}, \tag{2.50}$$

so that

$$\theta = \frac{\dot{A}}{A} + \frac{\dot{B}}{B} + \frac{\dot{C}}{C} = 3\frac{\dot{R}}{R}. \tag{2.51}$$

The three directional Hubble parameters are

$$H_1 = \frac{\dot{A}}{A}, \qquad H_2 = \frac{\dot{B}}{B}, \qquad H_3 = \frac{\dot{C}}{C}, \tag{2.52}$$

with mean value

$$H = \frac{1}{3}(H_1 + H_2 + H_3) = \frac{\dot{R}}{R}. \tag{2.53}$$

Since the spatial curvature vanishes, the general field equations of Section 2.1 simplify considerably. For an orthogonal perfect fluid the Hamiltonian constraint becomes

$$3H^2 = \sigma^2 + \rho, \tag{2.54}$$

the Raychaudhuri equation is

$$\dot{\theta} = -\frac{1}{3}\theta^2 - 2\sigma^2 - \frac{1}{2}(\rho + 3p), \tag{2.55}$$

and the shear propagation equations reduce to

$$\dot{\sigma}_{\alpha\beta} + \theta\,\sigma_{\alpha\beta} = 0. \tag{2.56}$$

Thus

$$\sigma_{\alpha\beta} = \Sigma_{\alpha\beta} R^{-3}, \tag{2.57}$$

where $\Sigma_{\alpha\beta}$ is a constant trace-free tensor. If we define

$$\Sigma^2 \equiv \frac{1}{2}\Sigma_{\alpha\beta}\Sigma^{\alpha\beta}, \tag{2.58}$$

then

$$\sigma^2 = \Sigma^2 R^{-6}. \tag{2.59}$$

This $R^{-6}$ behaviour of the shear is one of the key properties of Bianchi-I cosmology.

It is also convenient to write the metric functions in the form

$$A = R\,e^{\beta_1}, \qquad B = R\,e^{\beta_2}, \qquad C = R\,e^{\beta_3}, \qquad \beta_1 + \beta_2 + \beta_3 = 0. \tag{2.60}$$

Then

$$\dot{\beta}_i = \Sigma_i R^{-3}, \qquad \Sigma_1 + \Sigma_2 + \Sigma_3 = 0, \tag{2.61}$$

where the $\Sigma_i$ are constants related to the diagonal components of the shear.

***Vacuum case: the Kasner solution***

The vacuum Bianchi-I model is obtained by setting

$$\rho = p = 0.$$

Equation (2.54) then gives

$$3H^2 = \Sigma^2 R^{-6}, \tag{2.62}$$

so that $R^3 \propto t$. Choosing the origin of proper time at the singularity, one obtains

$$R \propto t^{1/3}. \tag{2.63}$$

Using Eq. (2.61), the directional scale factors are then power laws in $t$, and the metric takes the well-known Kasner form

$$ds^2 = -dt^2 + t^{2p_1}dx^2 + t^{2p_2}dy^2 + t^{2p_3}dz^2. \tag{2.64}$$

The exponents $p_i$ satisfy the two Kasner conditions

$$p_1 + p_2 + p_3 = 1, \tag{2.65}$$

$$p_1^2 + p_2^2 + p_3^2 = 1. \tag{2.66}$$

Conversely, every set of constants $(p_1, p_2, p_3)$ satisfying Eqs. (2.65)-(2.66) defines a vacuum Bianchi-I solution.

Except for the special flat cases

$$(1,0,0) \quad \text{and permutations,} \tag{2.67}$$

the Kasner spacetime possesses a genuine anisotropic curvature singularity at $t = 0$. Indeed, the Weyl curvature does not vanish, and the singularity is of a qualitatively different character from the Ricci singularity of FLRW cosmology.

The standard one-parameter representation is obtained by ordering the exponents as $p_1 \le p_2 \le p_3$ and introducing a parameter $u \ge 1$ such that

$$p_1 = -\frac{u}{1+u+u^2}, \qquad p_2 = \frac{1+u}{1+u+u^2}, \qquad p_3 = \frac{u(1+u)}{1+u+u^2}. \tag{2.68}$$

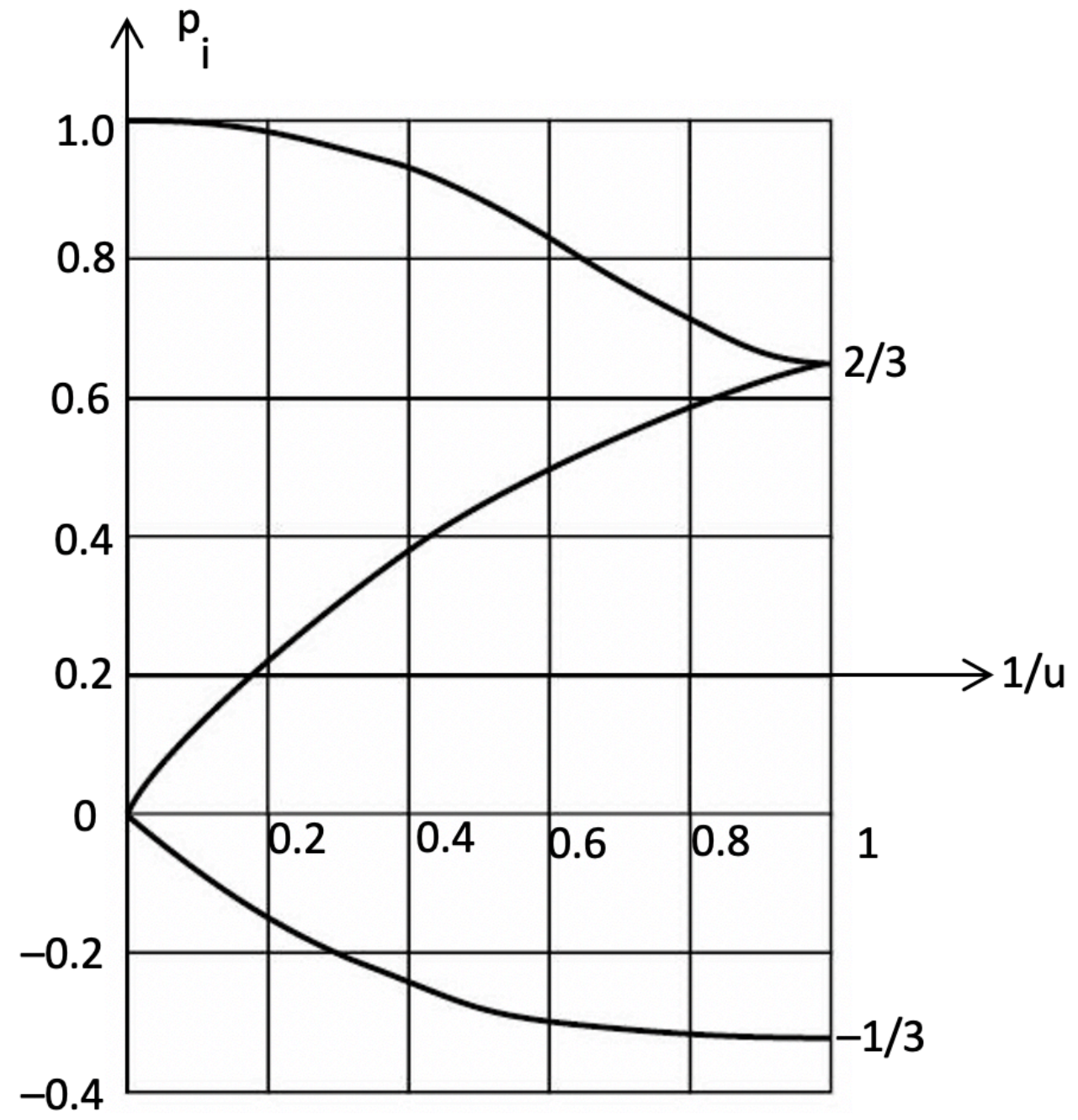

**Figure 3**: Kasner exponents $p_i$ as functions of 1/u, for u ≥ 1.

For the generic Kasner solution, one exponent is negative and the other two are positive. Thus, as $t \to 0^+$, two scale factors contract to zero while the third diverges. The singularity is then of the familiar cigar type. The degenerate cases with one exponent equal to unity and the other two equal to zero correspond to flat Minkowski spacetime written in nonstandard coordinates.

***Perfect-fluid Bianchi-I models***

Let us now consider a perfect fluid satisfying the barotropic equation of state

$$p = (\gamma - 1)\rho, \qquad 1 \le \gamma \le 2. \tag{2.69}$$

The conservation law gives

$$\rho = \rho_0 R^{-3\gamma}, \tag{2.70}$$

where $\rho_0$ is constant. Combining this with Eq. (2.59), the Hamiltonian constraint becomes

$$3H^2 = \Sigma^2 R^{-6} + \rho_0 R^{-3\gamma}. \tag{2.71}$$

This compact equation already contains the essential qualitative information.

For

$$1 \le \gamma < 2, \tag{2.72}$$

assuming $\Sigma^2>0$ (the isotropic limit $\Sigma^2=0$ being excluded from this shear-dominance argument), the matter density diverges as R^-3γ, whereas the shear diverges more rapidly as R^-6. Hence, near the singularity,

$$\Sigma^2 R^{-6} \gg \rho_0 R^{-3\gamma}, \qquad R \to 0, \tag{2.73}$$

and the dynamics becomes asymptotically vacuum-like. The solution therefore approaches a Kasner regime in the neighbourhood of the initial singularity.

This is the precise form of the often-stated result that, for ordinary perfect fluids with non-zero shear ($\Sigma^2>0$), anisotropy dominates matter near a Bianchi-I singularity; the isotropic limit $\Sigma^2=0$ is not part of this argument.

The case

$$\gamma = 2 \tag{2.74}$$

is exceptional. Then both terms in Eq. (2.71) scale in the same way,

$$\rho \sim R^{-6}, \qquad \sigma^2 \sim R^{-6}, \tag{2.75}$$

so that matter remains dynamically relevant all the way to the singularity. This is the *stiff-fluid* case, closely related to a massless scalar field, and it leads to the Jacobs family of exact Bianchi-I solutions rather than to vacuum Kasner asymptotics.

In the stiff-fluid case the metric may still be written in power-law form,

$$ds^2 = -dt^2 + t^{2p_1}dx^2 + t^{2p_2}dy^2 + t^{2p_3}dz^2, \tag{2.76}$$

but the exponents no longer satisfy the vacuum Kasner condition (2.66). Instead,

$$p_1 + p_2 + p_3 = 1, \tag{2.77}$$

while

$$p_1^2 + p_2^2 + p_3^2 < 1, \tag{2.78}$$

the deviation from unity being determined by the stiff-fluid density. Thus the stiff-fluid singularity is still anisotropic in general, but it is not vacuum-dominated.

This distinction is of considerable modern importance. In later work on cosmological singularities, stiff fluids and scalar fields form the basic examples of quiescent or non-oscillatory singular behaviour. We need only note the exception here; its wider significance will be taken up in Primer III (Andersson and Rendall, 2001).

***Geometrical character of the singularity***

The Bianchi-I solutions illustrate very clearly that the vanishing of the average scale factor $R$ does not by itself specify the detailed geometry of the singularity. Even within this simplest anisotropic class, the singular approach may differ from the isotropic FLRW case.

In vacuum or for $1 \le \gamma < 2$, the singularity is asymptotically Kasner-like and therefore generically cigar-shaped. In fluid models, depending on the relative behaviour of the directional scale factors, one may also encounter the point-, barrel-, pancake- or cigar-type singularities familiar from the classical classification of homogeneous cosmologies. What remains universal is that, except in the flat special cases, the past boundary is a genuine curvature singularity.

Bianchi I therefore serves as the simplest prototype of two ideas that will recur throughout this article. First, spatial homogeneity does not imply isotropy. Second, the presence of matter does not necessarily control the singular behaviour: for ordinary barotropic fluids on anisotropic branches with $\Sigma^2>0$, shear becomes asymptotically dominant, while the isotropic branch and stiff matter require separate treatment.

These conclusions motivate the more general singularity results of Section 3, where we move beyond the special solvable example of Bianchi I and ask which properties survive in arbitrary homogeneous cosmologies.

## 3. Singularities in Spatially Homogeneous Cosmologies

### *3.1. Focusing and Breakdown of the Homogeneous Cauchy Development*

The Bianchi-I models studied in Section 2.3 provide explicit examples of spatially homogeneous cosmologies possessing curvature singularities. Rather than examining each Bianchi type separately, we now seek conclusions that follow from spatial homogeneity and the Einstein equations without detailed reference to the algebraic structure of the homogeneity group.

Let $S = S(t_1)$ be a spacelike homogeneous hypersurface on which a three-dimensional isometry group acts simply transitively, and consider the maximal globally hyperbolic development $D(S)$ of homogeneous initial data prescribed on $S$. By uniqueness of the Cauchy development, the symmetries of the initial data extend throughout $D(S)$.

We assume that the matter content is a perfect fluid and that, with $\Lambda = 0$,

$$R_{ab} = (\rho + p)u_a u_b + \tfrac{1}{2}(\rho - p)g_{ab}, \qquad u^a u_a = -1. \tag{3.1}$$

We further impose the energy inequalities

$$\rho + p \geq 0, \qquad \rho + 3p \geq 0. \tag{3.2}$$

For a perfect fluid in four dimensions, the conjunction of the two inequalities in Eq. (3.2) is the strong energy condition relevant to the Raychaudhuri focusing argument.

Let $n^a$ be the geodesic unit normal to the homogeneous hypersurfaces. As in Section 2.2, the fluid velocity can be written

$$u^a = \cosh\beta\, n^a + \sinh\beta\, e^a,$$

where $e^a$ is a unit spatial vector tangent to $S(t)$. Hence

$$R_{ab}n^a n^b = \tfrac{1}{2}(\rho + 3p) + (\rho + p)\sinh^2\beta. \tag{3.3}$$

The normal congruence is geodesic and vorticity-free. Its Raychaudhuri equation therefore reads

$$\dot{\theta} = -\tfrac{1}{3}\theta^2 - 2\sigma^2 - \tfrac{1}{2}(\rho + 3p) - (\rho + p)\sinh^2\beta, \tag{3.4}$$

where the dot denotes differentiation along $n^a$.

Under the assumptions (3.2),

$$\dot{\theta} \leq -\tfrac{1}{3}\theta^2. \tag{3.5}$$

This elementary inequality already has a powerful consequence.

Suppose first that $\theta(t_1) = \theta_1 < 0$. Then

$$\frac{d}{dt}\left(\frac{1}{\theta}\right) = -\frac{\dot{\theta}}{\theta^2} \geq \frac{1}{3},$$

and therefore, as long as the denominator remains positive,

$$\theta(t) \leq \frac{\theta_1}{1+\frac{\theta_1}{3}(t-t_1)}, \qquad 1 + \frac{\theta_1}{3}(t - t_1) > 0. \tag{3.6}$$

This comparison shows that the regular contracting development cannot persist for a proper-time interval exceeding $3/|\theta_1|$. It provides an upper bound on the focusing interval; it does not identify the comparison blow-up time with the actual endpoint.

Conversely, if $\theta(t_1)>0$, the same argument applied toward the past gives the analogous finite-time focusing statement, with an interval not exceeding $3/\theta(t_1)$.

Finally, if $\theta(t_1) = 0$ and at least one of the non-negative terms $\sigma^2$, $\rho+3p$, or $(\rho+p)\sinh^2\beta$ is nonzero, Eq. (3.4) implies that $\theta$ immediately becomes negative toward the future and positive toward the past. Focusing therefore occurs in both time directions. The exceptional case in which all these terms vanish corresponds to a locally flat or otherwise degenerate non-focusing situation and must be treated separately.

A familiar example of the former situation is a recollapsing closed FLRW universe at its moment of maximum expansion.

***Collapse of the homogeneous volume***

Let $V(t)$ denote the local volume element of the homogeneous hypersurfaces and define the average linear scale

$$\ell(t) \equiv V^{1/3}(t). \tag{3.7}$$

Since $\theta = 3(d\ell/dt)/\ell$, the Raychaudhuri equation gives

$$3\frac{\ddot{\ell}}{\ell} = \dot{\theta} + \tfrac{1}{3}\theta^2 = -2\sigma^2 - \tfrac{1}{2}(\rho + 3p) - (\rho + p)\sinh^2\beta \leq 0. \tag{3.8}$$

Thus $\ell(t)$ is concave. If $\theta_1<0$, then $d\ell/dt$ at $t=t_1$ equals $(\theta_1/3)\ell(t_1)<0$, and concavity implies

$$\ell(t) \le \ell(t_1) + \dot{\ell}(t_1)(t - t_1).$$

If the homogeneous solution remains regular up to the focusing endpoint, the right-hand side reaches zero no later than $t_1+3/|\theta_1|$. Hence

$$V = \ell^3 \to 0. \tag{3.9}$$

If instead the maximal homogeneous development terminates earlier, that earlier endpoint must be analysed separately as a curvature or extendibility boundary. In the matter-singularity results below we explicitly consider focusing branches for which $V\to 0$.

On a $V\to 0$ focusing branch, the determinant of the induced three-metric tends to zero and the Gaussian normal foliation by regular spacelike homogeneous hypersurfaces degenerates at the finite-time boundary.

The essential conclusion is therefore that the Raychaudhuri inequality forces finite-time focusing, while the full homogeneous evolution determines whether the corresponding branch reaches $V\to 0$ or terminates earlier at another boundary. The nature of that boundary must be established separately.

***What does the focusing boundary represent?***

On the $V\to 0$ focusing branch assumed in the matter-singularity theorem below, the focusing argument by itself does not determine the nature of the boundary at $t_0$. In particular, the vanishing of the spatial volume does not by itself prove that a scalar curvature invariant diverges.

Several possibilities must therefore be distinguished.

The boundary may be a genuine curvature singularity, in which case curvature components or scalar polynomial invariants become unbounded. This is what happens, for example, in the generic Kasner solutions discussed in Section 2.3.

Alternatively, the homogeneous foliation may degenerate at a null boundary or Cauchy horizon through which the spacetime admits some form of extension. The Taub region of Taub-NUT spacetime provides the classical vacuum example, and related phenomena will appear later in tilted perfect-fluid models.

Consequently, the appropriate conclusion of the Raychaudhuri argument is not that spatial homogeneity itself must necessarily "break down," but rather that the *regular homogeneous Cauchy development reaches a finite-time boundary*. Determining whether this boundary is a matter singularity, a Weyl-curvature singularity, a non-scalar curvature singularity, or an extendible Cauchy horizon requires additional information from the field equations.

This question is the subject of the following sections.

### *3.2. Matter Singularities*

The focusing argument of the preceding section establishes that the regular homogeneous Cauchy development reaches a finite-proper-time boundary, but does not by itself determine whether this boundary is a curvature singularity. In many important cases, however, one can prove that the matter density diverges there and hence that the boundary is a scalar Ricci curvature singularity.

For a perfect fluid with $\Lambda = 0$,

$$R_{ab}R^{ab} = \rho^2 + 3p^2. \tag{3.10}$$

Consequently,

$$\rho \to \infty$$

necessarily implies

$$R_{ab}R^{ab} \to \infty. \tag{3.11}$$

We shall refer to such a singularity as a 'matter singularity'. This terminology emphasizes that the divergence is already present in the Ricci curvature determined locally by the stress-energy tensor, independently of any possible divergence of the Weyl tensor.

In the matter-singularity results that follow we exclude the vacuum case and assume, on the regular branch, ϱ>0 and ϱ+p>0, together with a sufficiently regular barotropic equation of state p=p(ϱ). These assumptions ensure that the functions r(ϱ) and s(ϱ) introduced in Eq. (2.37) are well defined. Additional pressure bounds are stated explicitly where required.

***Orthogonal perfect-fluid models***

The simplest case is that of an orthogonal perfect fluid, $u^a = n^a$. Then the expansion of the fluid coincides with the expansion $\theta$ of the normal congruence, and the conservation equation is

$$\dot{\rho} + (\rho + p)\theta = 0. \tag{3.12}$$

Recall the function $s(\rho)$ introduced in Eq. (2.37),

$$\frac{ds}{s} = \frac{d\rho}{\rho+p}. \tag{3.13}$$

Equation (3.12) immediately gives

$$\frac{\dot{s}}{s} = -\theta.$$

Since $\theta = \dot{V}/V$, where $V$ is the volume element of a homogeneous hypersurface,

$$sV = M = \text{const.} \tag{3.14}$$

At the focusing boundary established in Section 3.1, $V \to 0$, and therefore

$$s \to \infty. \tag{3.15}$$

To infer the behaviour of $\rho$, we impose the same physically standard bounds as in Eq. (1.10),

$$-\frac{\rho}{3} \le p \le \rho. \tag{3.16}$$

They imply

$$\frac{1}{2}\frac{d\rho}{\rho} \le \frac{d\rho}{\rho + p} \le \frac{3}{2}\frac{d\rho}{\rho},$$

and hence, up to positive integration constants,

$$C_1\rho^{1/2} \le s \le C_2\rho^{3/2}. \tag{3.17}$$

Thus $s \to \infty$ implies $\rho \to \infty$, and Eq. (3.10) shows that a scalar Ricci curvature invariant diverges.

We may summarize the result as follows.

Theorem 3.1. *Consider a spatially homogeneous cosmology containing a non-vacuum orthogonal perfect fluid with positive density, a regular barotropic equation of state, ρ+p>0, and satisfying Eq. (3.16). If V→0 along the focusing branch, then ρ→∞ and the Ricci-squared invariant of Eq. (3.10) diverges; the boundary is therefore a matter singularity.*

The geometrical shape of the singularity need not be isotropic. As already illustrated by Bianchi I, the directional scale factors may display point-, barrel-, pancake- or cigar-type behaviour. The theorem establishes curvature blow-up, but does not by itself determine the detailed asymptotic dynamics. That problem will be addressed in *Primer III*.

***Kantowski-Sachs models***

The preceding analysis can be extended to the Kantowski-Sachs cosmologies introduced in Section 1.1. These models are spatially homogeneous but do not admit a three-dimensional simply transitive subgroup of the type used in the Bianchi classification. Their metric may be written

$$ds^2 = -dt^2 + A^2(t)\,dr^2 + B^2(t)(d\vartheta^2 + \sin^2\vartheta\,d\varphi^2). \tag{3.18}$$

The homogeneous spatial volume element is proportional to

$$V = AB^2, \quad (3.19)$$

and the expansion scalar is therefore

$$\theta = \frac{\dot{A}}{A} + 2\frac{\dot{B}}{B} = \frac{\dot{V}}{V}. \quad (3.20)$$

For a perfect fluid with $\rho + p \neq 0$, the mixed Einstein equation $G_{0r} = 0$ forces the fluid four-velocity to be orthogonal to the homogeneous hypersurfaces. The Raychaudhuri equation consequently takes the same form as for an orthogonal Bianchi model,

$$\dot{\theta} = -\frac{1}{3}\theta^2 - 2\sigma^2 - \frac{1}{2}(\rho + 3p). \quad (3.21)$$

Under the strong energy condition, a nontrivial expanding or contracting branch therefore focuses within finite proper time exactly as in Section 3.1. Since the fluid is orthogonal, Eq. (3.14) again applies. With the bounds (3.16), the vanishing of $V$ forces $\rho \to \infty$.

Hence:

Theorem 3.2. *A nonstatic Kantowski-Sachs cosmology containing a non-vacuum orthogonal perfect fluid satisfying the same regularity and pressure assumptions as Theorem 3.1 develops a matter singularity along any focusing branch for which V→0.*

This formulation makes explicit the assumptions that were left implicit in some of the early statements of the theorem.

***Tilted perfect-fluid models***

The situation is less immediate when the fluid is tilted. The reason is already apparent from Section 2.2: the homogeneous volume may collapse while the Lorentz factor

$$\Gamma = \cosh\beta$$

between the fluid and the normal congruences becomes arbitrarily large. In principle, extreme tilt might therefore allow the density measured in the fluid rest frame to remain finite even as the normal congruence focuses.

The classical analysis of Ellis and King shows that this possibility is strongly restricted. In the notation adopted here, the unit spatial tilt direction and the Class-B commutator vector are both expressed in the same symmetry-adapted orthonormal frame associated with the normal congruence. This specification is important because the boundedness hypothesis below is a statement about their components in that common frame.

Theorem 3.3 (Ellis and King, 1974). *Consider a spatially homogeneous tilted non-vacuum perfect-fluid model with positive density, a regular barotropic equation of state, p≥0, and satisfying the energy conditions of Section 3.1. If the contraction of the Class-B commutator vector with the unit tilt direction is bounded from below along the focusing branch, then that branch terminates at a matter singularity: ρ→∞ and the Ricci-squared invariant of Eq. (3.10) diverges.*

The proof is more involved than in the orthogonal case, but its essential mechanism can be summarized compactly. Introduce the two functions already defined in Eq. (2.37),

$$\frac{dr}{r} = \frac{dp}{\rho+p}, \qquad \frac{ds}{s} = \frac{d\rho}{\rho+p}. \quad (3.22)$$

They satisfy

$$\frac{d(rs)}{rs} = \frac{d(\rho+p)}{\rho+p},$$

so that their normalization may be chosen such that

$$rs = \rho + p. \quad (3.23)$$

This is the corrected form of the normalization relation used in the original Ellis–King argument.

The fluid conservation equations relate $s$, the collapsing homogeneous volume, and the tilt factor $\cosh\beta$. If $\rho$ were to remain bounded while $V \to 0$, the only possible compensation would be

$$|\beta| \to \infty, \quad (3.24)$$

that is, extreme tilt.

The Hamiltonian constraint, Raychaudhuri equation and tilt evolution equation then place increasingly strong restrictions on this possibility. For the relevant non-IX Bianchi types, one convenient part of the classical argument uses the non-positivity of the spatial scalar curvature together with the assumed lower bound on the contraction just specified to control the growth of the tilt. This shows that extreme tilt cannot compensate sufficiently rapidly for the collapse of the homogeneous volume if the density is to remain bounded. Since the present proof sketch does not cover Bianchi IX, we do not include type IX in the corollary below; its singular behaviour is discussed separately in the subsequent paragraph on rigorous Bianchi VIII/IX results.

For Class A, the commutator vector vanishes identically, so the boundedness condition is automatically satisfied. For the non-IX Class-A types covered by the preceding argument this gives the following corollary:

Corollary. Every tilted non-vacuum perfect-fluid Bianchi model of Class A and type I, II, $VI_0$, $VII_0$ or VIII satisfying the hypotheses of Theorem 3.3 and the assumptions of the preceding argument possesses a matter singularity along a $V \to 0$ focusing branch.

Thus, under these assumptions, the non-IX Class-A types I, II, $VI_0$, $VII_0$ and VIII lead to a Ricci curvature singularity analogous, in this restricted sense, to the Big-Bang singularity of FLRW cosmology. Bianchi IX is kept separate here because the non-positive-spatial-curvature step used above does not apply to it. The Class-B models that evade the hypothesis of Theorem 3.3 are precisely where more subtle possibilities arise. These lead us in Section 3.3 to intermediate or whimper singularities.

***Vacuum models: curvature blow-up without a matter singularity***

It is important to distinguish a *matter singularity* from a *curvature singularity*. The former obviously cannot occur in vacuum, but vacuum spacetimes may nevertheless possess strong Weyl-curvature singularities. The Kasner solution of Section 2.3 already provides the simplest example.

The classical literature also exhibited exceptional vacuum solutions in which the focusing of the homogeneous foliation does not correspond to scalar curvature blow-up. The best-known example is the Taub region of the Taub-NUT solution, whose finite-time boundary is a Cauchy horizon across which the spacetime can be extended. Such examples helped motivate the distinction, emphasized in Section 3.1, between geodesic focusing and curvature singularity.

The modern picture is considerably sharper. Ringström proved that the maximal globally hyperbolic developments of non-Taub-NUT Bianchi IX vacuum data and non-NUT Bianchi VIII vacuum data are $C^2$-inextendible and that a curvature invariant becomes unbounded in the incomplete direction of inextendible causal geodesics (Ringström, 2000). Thus the extendible Taub/NUT cases are exceptional rather than representative of the general VIII/IX vacuum behaviour.

Ringström subsequently considered Class-A models containing an orthogonal perfect fluid with a linear equation of state characterized by a constant parameter $\gamma$ and with $\Lambda=0$. The paper assumes $2/3<\gamma\leq 2$; its global Class-A curvature statements are formulated for $1\leq\gamma\leq 2$. In the non-stiff Bianchi-IX case ($\gamma<2$), generic solutions approach the Bianchi-IX attractor formed by the vacuum type-I and type-II subsets, whereas the stiff-fluid case $\gamma=2$ is qualitatively different and converges to a non-vacuum type-I state. Ringström also established curvature blow-up along incomplete causal geodesics for the Class-A developments covered by the theorem, apart from the exceptional vacuum Taub cases (Ringström, 2001). We postpone the dynamical content of these results - Kasner states, Bianchi-II transitions and the Mixmaster attractor - to Primer III.

The conclusion relevant here is simpler: finite-time focusing in homogeneous cosmology very often corresponds to genuine curvature blow-up, but the mechanism need not be Ricci blow-up and exceptional extendible Cauchy horizons do exist. This distinction prepares the ground for the intermediate singularities considered next.

### *3.3. Intermediate, Whimper, and Kinematic Singularities*

The matter-singularity results of Section 3.2 cover all Class-A models satisfying the stated hypotheses and a large family of Class-B models. Class B nevertheless admits a qualitatively different possibility. The homogeneous volume may collapse and the normal congruence may reach the boundary of its Cauchy development while the fluid density and all scalar polynomial curvature invariants remain finite.

This phenomenon was uncovered in the late 1960s and developed systematically by Ellis and King in their aptly titled paper *Was the Big Bang a Whimper?* (1974).

***The classical Ellis–King whimper***

The prototype is the locally rotationally symmetric tilted Bianchi-V solution originally obtained by Farnsworth and analyzed globally by Shepley. In this model the matter is dust,

$$p = 0,$$

The homogeneous Cauchy development $D^+(S)$ reaches an associated future Cauchy horizon $H^+(S)$, beyond which the initial data on S no longer determine the spacetime uniquely. The energy density remains finite there and no scalar polynomial curvature invariant need diverge.

It is important to separate this causal statement from the curvature statement. A Cauchy horizon by itself is not a curvature singularity. In the classical Ellis-King whimper solutions, however, the associated intermediate boundary is non-scalar in the stricter sense that curvature or matter components become unbounded in a parallelly propagated frame along an inextendible causal geodesic that reaches the boundary at finite affine parameter, while scalar invariants remain finite. In the standard construction the relevant incomplete curves may be taken to be null geodesics. The parallelly propagated frame becomes infinitely boosted relative to the group-invariant frame. It is this parallelly propagated blow-up, rather than the existence of the Cauchy horizon alone, that justifies the term non-scalar curvature singularity.

The essential causal distinction is

$$J^+(S) \neq D^+(S) \qquad (3.25)$$

more precisely,

$$J^+(S) - D^+(S) \neq \emptyset \qquad (3.26)$$

Beyond the Cauchy horizon there are spacetime regions whose properties are no longer uniquely determined by the initial data on S.

In the Farnsworth–Shepley dust solution, the fluid worldlines themselves can be continued through this horizon and subsequently terminate at an ordinary Ricci curvature singularity where $\varrho \to \infty$. Thus the whimper is an intermediate singularity in a very literal sense: it separates the original homogeneous Cauchy development from a further region in which the matter may eventually encounter a conventional "bang."

Ellis and King showed that this behaviour is not peculiar to one exact solution. In the regular tilted perfect-fluid setting used in their analysis, their Class-B existence and Cauchy-horizon results may be restated in the present notation as follows; the theorem numbering below is local to this review (Ellis and King, 1974).

**Theorem 3.4 (Ellis and King, 1974).** *For every Bianchi type of Class B, there exist tilted perfect-fluid cosmologies for which $J^+(S) - D^+(S) \neq \emptyset$. The timelike geodesics normal to S are incomplete within $D^+(S)$ and reach its future boundary after finite proper time.*

Theorem 3.5 (Ellis and King, 1974). *Under the same classical tilted-perfect-fluid assumptions, when such a boundary occurs, the null generators of the Cauchy horizon $H^+(S)$ terminate at the associated intermediate boundary, while the geodesics normal to the original homogeneous hypersurface remain inside $D^+(S)$ and end at that boundary. The causal-horizon statement and any curvature blow-up criterion must be assessed separately.*

The important feature is therefore that geodesic incompleteness, parallelly propagated curvature behaviour and loss of predictability need not coincide with divergence of scalar invariants. The normal congruence reaches a finite-time boundary even though the usual scalar indicators of a matter singularity remain innocuous.

This is precisely the kind of phenomenon for which the broader singularity classification developed in *Primer I* becomes necessary: scalar polynomial invariants alone do not exhaust the possible pathological behaviours of spacetime.

***Extreme tilt as the underlying mechanism***

The relation with the tilted-fluid discussion of Section 2.2 is immediate. Avoidance of the matter-singularity mechanism of Theorem 3.3 requires, in the relevant Class-B solutions, that the relative Lorentz factor

$$\Gamma = \cosh\beta$$

become arbitrarily large. Thus

$$|\beta| \to \infty, \qquad |v| = \tanh|\beta| \to 1. \tag{3.27}$$

The fluid worldlines consequently become asymptotically null relative to the observers orthogonal to the homogeneous hypersurfaces.

This is the central physical mechanism behind the whimper: the metric and the matter density can remain comparatively mild while the relative motion of matter with respect to the homogeneous foliation becomes singular.

The significance of this observation became much clearer with the later dynamical-systems treatment of tilted Bianchi cosmologies.

***Kinematic singularities***

A particularly illuminating modern development was provided by Coley, Hervik, Lim and MacCallum in 2009. They reconsidered LRS tilted perfect-fluid Bianchi-V cosmologies and identified what they termed *kinematic singularities* (Coley et al., 2009).

The phenomenon is remarkable because the spacetime itself can be future geodesically complete: future-directed causal geodesics can be extended indefinitely, and observers orthogonal to the homogeneous hypersurfaces encounter no spacetime singularity in that direction. Nevertheless, the accelerated tilted-fluid congruence becomes inextendible after a finite amount of its own proper time.

The limit is again one of extreme tilt,

$$|v| \to 1,$$

but now some kinematic quantities measured by the fluid observers, notably the fluid Hubble expansion, diverge:

$$\hat{H} \to \infty$$

in finite fluid proper time, while the matter density may tend to zero rather than infinity.

In the strict terminology proposed by Coley et al., a kinematic singularity is therefore characterized by finite fluid proper time, divergence of one or more fluid kinematic variables, but bounded Ricci and Weyl curvature components.

This is not a spacetime singularity in the usual geodesic-incompleteness sense. The pathology belongs to the accelerated fluid congruence rather than to the underlying Lorentzian manifold.

Coley et al. obtained an even stronger result. For any prescribed positive integer $N$, there exist LRS Bianchi-V examples for which not only the curvature tensor but also its covariant derivatives through order $N$ remain bounded at the kinematic singularity. At sufficiently high derivative order a divergence may nevertheless appear.

This provides an instructive refinement of the original Ellis-King picture. A kinematic singularity need not imply a singularity of the Lorentzian spacetime: one may instead have a spacetime that is regular and future geodesically complete in the direction under consideration, while a physically distinguished accelerated matter congruence becomes singular.

It is therefore useful to distinguish four notions:

• matter singularity: ϱ and Ricci scalar invariants diverge;

• parallelly propagated curvature singularity: curvature components diverge in a parallelly propagated frame along an inextendible causal geodesic that reaches the boundary at finite affine parameter, although scalar invariants may remain finite;

• Cauchy-horizon boundary: the original Cauchy data cease to determine the extension uniquely; this causal property alone does not imply curvature blow-up;

• kinematic singularity: the fluid congruence becomes singular in finite fluid proper time although the spacetime may remain future geodesically complete and curvature components bounded.

The boundaries between these categories depend on which geometrical or physical structure is being regarded as fundamental: the spacetime metric, a preferred congruence, or the complete cosmological model (M, g, u).

***The return of the whimper: dipole cosmology***

Remarkably, the Ellis–King whimper has recently reappeared in a quite different cosmological context. Allahyari, Ebrahimian, Mondol and Sheikh-Jabbari investigated in detail the Big-Bang limit of *dipole cosmology*, an axisymmetric tilted Bianchi-V framework designed to allow a preferred cosmological direction. Their analysis explicitly builds on the King–Ellis tilted-cosmology formalism (Allahyari et al., 2025).

For a single perfect fluid with constant equation-of-state parameter, they find two qualitatively distinct generic approaches to the limit in which the overall scale factor tends to zero.

For one sign of the shear, the tilt becomes small and the universe approaches a strongly anisotropic, shear-dominated curvature singularity. Matter density and curvature invariants diverge.

For the opposite sign, however,

$$|\beta| \to \infty$$

while

$$\rho \to \rho_0 < \infty. \qquad (3.28)$$

All scalar polynomial curvature invariants remain finite. Allahyari et al. explicitly identify this branch as an Ellis–King whimper singularity.

The geometry near this branch is highly anisotropic and pancake-like. Allahyari et al. evaluate the Riemann tensor in a tetrad basis of the homogeneous metric and, in their Weyl decomposition, choose the metric-comoving timelike observer v^μ = (1, 0, 0, 0), rather than the tilted fluid four-velocity. Although all scalar polynomial curvature invariants remain finite, certain Riemann components in that tetrad diverge like $H^2$ and enter the geodesic-deviation equation. This establishes strong observer-frame tidal behaviour. However, their calculation does not by itself show that the chosen tetrad is parallelly propagated along an inextendible causal geodesic reaching the limiting surface at finite affine parameter; accordingly, we do not identify this frame-component blow-up automatically with a parallelly propagated curvature singularity.

This distinction deserves emphasis. Allahyari et al. relate their Big-Bang whimper to the modern terminology of kinematic singularities, but it is not identical to the strict Coley-Hervik-Lim-MacCallum case. In the latter, the spacetime is future geodesically complete and the Ricci and Weyl curvature components remain bounded while the fluid congruence becomes singular. In the former, scalar invariants remain finite but selected tetrad components of the Riemann tensor diverge. The two phenomena are linked by extreme tilt, but we do not identify the latter automatically with a parallelly propagated curvature singularity.

The same 2025 analysis extends the result to a tilted multi-fluid $\Lambda$CDM-like model. Depending on the initial signs of the shear and the fluid tilts, the past limit is again either a shear-dominated curvature singularity or a whimper-type boundary with extreme tilt and finite scalar invariants.

Thus an effect first discovered in a rather specialized exact Bianchi-V solution more than fifty years ago has acquired renewed relevance in contemporary tilted cosmology.

***A modern interpretation***

The successive developments from Ellis and King to Coley et al. and Allahyari et al. reveal an important hierarchy.

A cosmological boundary need not be characterized by $\varrho \to \infty$ or by the divergence of a scalar polynomial curvature invariant. It may instead involve parallelly propagated curvature blow-up, loss of global hyperbolicity at a Cauchy horizon, frame-dependent tidal blow-up, or solely the singular behaviour of an accelerated matter congruence. These possibilities should be distinguished rather than grouped under a single criterion.

Extreme tilt provides a particularly natural mechanism for separating these notions. As

$$|v| \to 1,$$

the fluid and geometrically preferred normal frames become infinitely boosted relative to one another. Quantities that remain regular for one family of observers may then become unbounded for the other.

The old question "Was the Big Bang a whimper?" therefore remains meaningful, although its modern answer is more nuanced than the one envisaged in 1974. A vanishing cosmological scale need not imply a matter singularity, and even the notion of singularity may depend crucially on whether one examines scalar curvature, parallelly propagated curvature, global causal structure, or the kinematics of the cosmological fluid.

These issues naturally lead to the question of what happens beyond an intermediate boundary and to the possible global evolutions of homogeneous cosmologies, to which we now turn.

### *3.4. Global Evolution and the Ellis-King Classification*

The preceding sections have described two qualitatively different boundaries of a spatially homogeneous Cauchy development. The first is a scalar curvature singularity, frequently a matter singularity with $\varrho \to \infty$. The second is an intermediate or whimper boundary associated with a Cauchy horizon, across which the original homogeneous initial data no longer determine the spacetime uniquely; whether this boundary is also a non-scalar curvature singularity is a separate geometric question.

The natural question is whether the evolution can nevertheless be followed beyond such a Cauchy horizon and, if so, what global structures may result.

In the Farnsworth-Shepley example discussed above, the fluid worldlines cross the horizon associated with the whimper and eventually terminate at a matter singularity. To generalize this result, Ellis and King introduced an additional assumption ensuring that no arbitrary new information is supplied beyond the Cauchy horizon. Their analysis provides one of the earliest systematic attempts to classify the complete global evolution of tilted homogeneous cosmologies (Ellis and King, 1974).

For clarity, the assumptions entering the global Ellis-King results should be kept distinct. In addition to the perfect-fluid Einstein equations and the energy and regularity assumptions used above, the original homogeneous Cauchy development is taken to be maximal and the homogeneous initial hypersurface S simply connected. No arbitrary new data are admitted across a Cauchy horizon. This is implemented in either of two ways: (i) the spacetime symmetry is required to continue beyond the horizon, with the isometry group acting simply transitively on three-dimensional group orbits, at least one of which is spacelike; or (ii), for an analytic equation of state $p = p(\varrho)$, one takes a maximal locally analytic extension of the original solution. In the classical Ellis-King treatment these continuation prescriptions are equivalent subject to the absence of a quasi-regular obstruction. Theorems 3.6 and 3.7 below are quoted only under these hypotheses (Ellis and King, 1974).

These assumptions are stronger than those required for the local singularity results of the preceding sections. In particular, preservation of spatial homogeneity across a horizon, analyticity of an extension, and geodesic completeness are logically distinct statements; none should be inferred from the others without the hypotheses just listed.

**The two global theorems**

With the time orientation chosen so that the focusing or whimper boundary lies toward the future, Ellis and King established the following result.

Theorem 3.6 (Ellis and King, 1974). *Under the global assumptions just listed, if the Bianchi type is not IX, every causal geodesic that is inextendible toward the opposite asymptotic time direction is complete. With t denoting proper time along the homogeneous normal congruence, as t tends to that asymptotic end the matter becomes dilute,*

$$\rho \to 0, \quad p \to 0, \tag{3.29}$$

the expansion and shear associated with the homogeneous normal congruence tend to zero, and the corresponding spatial length scale tends to infinity.

Thus, outside the recollapsing type-IX situation, the classical models considered by Ellis and King possess one asymptotically dilute and geodesically complete end. This statement concerns the geometrically preferred normal congruence and its proper time; it should not be transferred without qualification to the tilted fluid congruence. In particular, in the future-complete LRS Bianchi-V examples of Coley et al. with $4/3<\gamma<2$, the normal Hubble scalar tends to zero while the fluid Hubble scalar diverges.

The second theorem concerns models whose homogeneous Cauchy development ends at a whimper.

**Theorem 3.7 (Ellis and King, 1974).** *Under the same global assumptions, suppose that*

$$J^+(S) - D^+(S) \neq \emptyset \tag{3.30}$$

*Then the spacetime contains a spatially homogeneous region $D_1 = D(S)$, separated by a Cauchy horizon from an extended region $D_2$. The latter is stationary and spatially inhomogeneous and has one of three possible future behaviours:*

*a) the fluid worldlines terminate after finite proper time at a scalar curvature singularity;*

*b) the fluid worldlines continue indefinitely without encountering another singularity;*

*c) the spacetime crosses a second Cauchy horizon into another spatially homogeneous region $D_3$, whose time orientation is reversed relative to $D_1$.*

Cases b and c therefore allow the matter to cross one or two whimper boundaries without ever encountering a matter singularity. In case a, by contrast, the whimper is genuinely intermediate: after crossing it the fluid ultimately reaches an ordinary scalar curvature singularity.

The theorem itself does not require this final scalar singularity to be a matter singularity. This distinction was investigated in detail by Collins for LRS tilted Bianchi-V models with

$$p = (\gamma - 1)\varrho. \qquad (3.31)$$

For dust, the terminal scalar singularity is a matter singularity with $\rho \to \infty$. Collins found, however, that for some equations of state with $1 < \gamma \le 4/3$, the density remains finite even though a scalar curvature singularity occurs. In the terminology of the period this was called a *conformal singularity*; in modern language, the divergence cannot arise from the Ricci invariants determined by the bounded perfect-fluid variables and must instead involve the Weyl sector of the curvature. Collins also emphasized the striking sensitivity of the singular behaviour to the equation of state (Collins, 1974).

**The classical fivefold synthesis**

Within these assumptions, Ellis and King organized the possible global histories into five schematic classes, represented in Figure 4.

**(a) Bang-bang.** The spacetime remains spatially homogeneous between two scalar matter singularities: it emerges from a Big-Bang-type boundary and ends at a Big-Crunch-type boundary. The closed recollapsing FLRW model is the elementary example; classical recollapsing Bianchi-IX perfect-fluid solutions provide anisotropic analogues.

**(b) Bang.** One time direction ends at a scalar singularity, while the other is geodesically complete and expands indefinitely into an asymptotically dilute state, with $\rho, p \to 0$. Spatially flat and negatively curved FLRW models provide the isotropic prototypes, while many orthogonal and Class-A Bianchi models realize the same qualitative structure under the assumptions used here.

**(c) Bang-whimper.** One end contains a scalar curvature singularity, but the other side of the global spacetime includes a whimper/Cauchy-horizon boundary separating a homogeneous region from a stationary inhomogeneous extension, which in turn approaches an asymptotically dilute complete end. Tilted LRS Bianchi-V solutions provide classical examples of this behaviour.

**(d) Whimper.** Both asymptotic directions are dilute and geodesically complete, but the evolution passes through one intermediate whimper/Cauchy-horizon boundary. No scalar matter singularity is encountered. Ellis and King did not know an explicit solution realizing the complete global configuration.

**(e) Whimper-whimper.** The most elaborate possibility contains two homogeneous asymptotic regions, separated by a stationary inhomogeneous region and two Cauchy horizons carrying intermediate singularities. Again the density and pressure may remain finite throughout. No explicit realization of this complete global pattern was known in the original analysis.

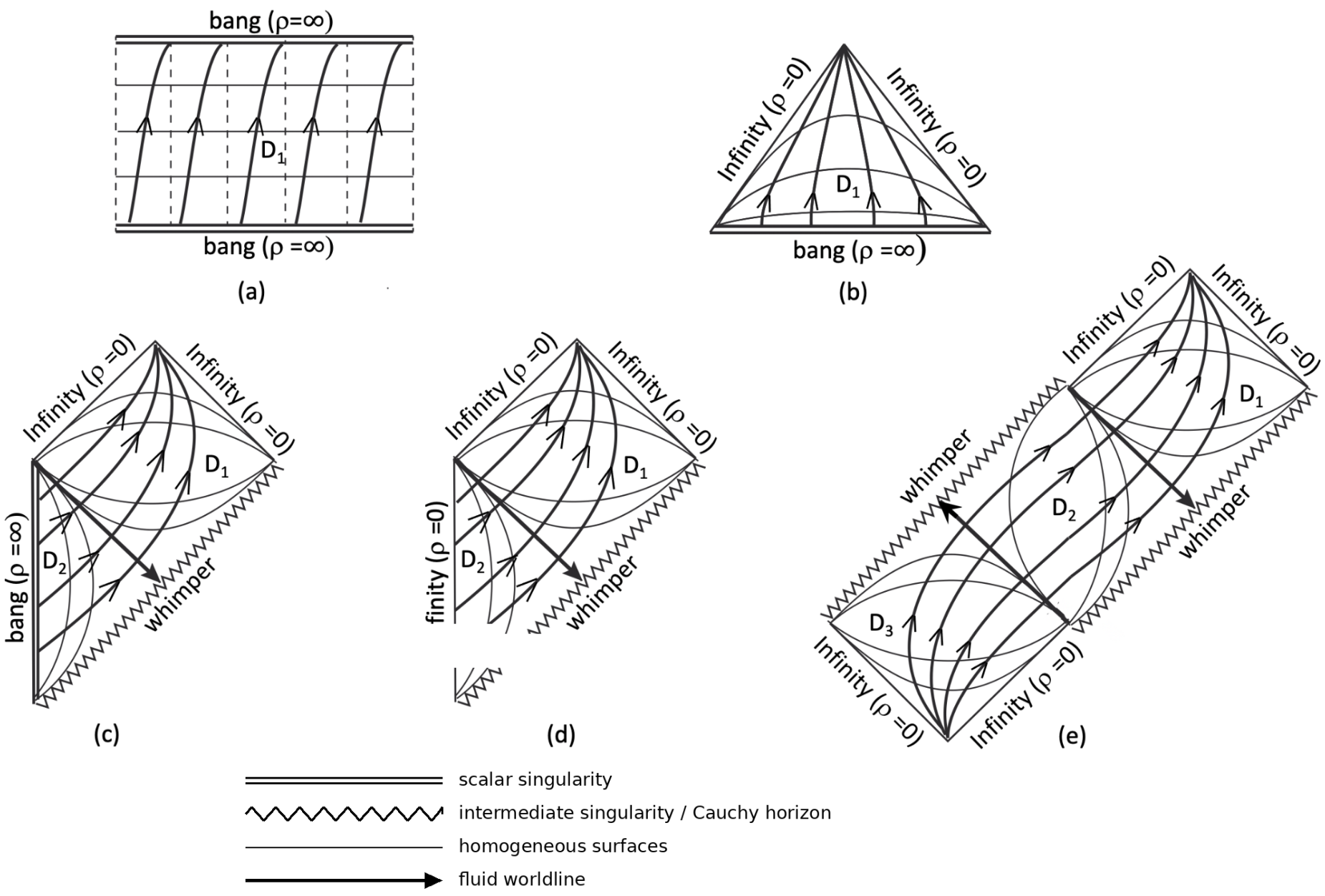

Figure 4. Classical Ellis-King classification of the possible global evolutions of spatially homogeneous perfect-fluid cosmologies under the assumptions described in the text: (a) bang-bang; (b) bang; (c) bang-whimper; (d) whimper; (e) whimper-whimper. Double lines represent scalar curvature singularities; wavy lines intermediate/whimper boundaries associated with Cauchy horizons; thin transverse curves homogeneous spatial hypersurfaces; and thick arrowed curves fluid worldlines.

**Stability and later perspective**

Already shortly after the Ellis-King analysis it became clear that whimper singularities are delicate objects. King showed that intermediate singularities are unstable under relevant perturbations, and Siklos subsequently investigated systematically the conditions under which whimper singularities can occur (King, 1975; Siklos, 1978).

This fragility is physically significant. The classical whimper configurations depend on a special relation between the geometry, the homogeneous group orbits and the tilted matter flow. Under perturbation, the intermediate boundary may be replaced by a scalar curvature singularity. In this sense, the classical question gradually shifted from the mere existence of whimper solutions to their genericity and stability.

The modern developments discussed in Section 3.3 reinforce rather than invalidate this conclusion. The kinematic singularities of Coley et al. (2009) and the whimper branch recently recovered in dipole cosmology by Allahyari et al. (2025) show that extreme-tilt singular behaviour is not merely an historical curiosity. However, these modern examples establish local or asymptotic whimper behaviour; they should not automatically be identified with the complete global “whimper” or “whimper-whimper” histories of panels (d) and (e). The 2025 dipole analysis, for example, demonstrates a Big-Bang whimper with divergent tilt and finite scalar invariants, but does not by itself realize the entire two-ended Ellis-King global diagram.

Thus the fivefold classification retains considerable conceptual value, but its proper status is now clear: it is a classical global synthesis for a restricted family of analytic perfect-fluid homogeneous cosmologies, rather than a universal classification of cosmological singularities.

Its enduring lesson is broader. Even within highly symmetric solutions of General Relativity, the global alternatives are richer than a simple dichotomy between a Big Bang and geodesically complete expansion. Scalar curvature blow-up, non-scalar singularities, Cauchy horizons, extreme tilt, and complete asymptotic ends can occur in different combinations.

This completes our analysis of the existence and nature of singularities in spatially homogeneous cosmologies. The much deeper question of their dynamical approach - Kasner epochs, Hamiltonian evolution, oscillatory Bianchi VIII and IX behaviour, Mixmaster dynamics and the BKL conjecture - will be the subject of *Primer III*.

## Conclusion

Spatially homogeneous cosmologies occupy a special position in the study of relativistic singularities. They retain enough symmetry to make the Einstein equations tractable, while allowing phenomena - anisotropy, shear, tilt, vorticity and nontrivial spatial curvature - that are absent from the FLRW models. They therefore provide the simplest setting in which the abstract concepts developed in *Primer I* can be confronted with explicit spacetime geometries.

The first general lesson is that finite-time focusing is extremely robust. For an initially contracting geodesic and hypersurface-orthogonal normal congruence satisfying the stated energy conditions, the Raychaudhuri inequality bounds the duration of a regular contracting development. Branches reaching $V\to 0$ undergo volume collapse; earlier endpoints require separate analysis. In broad classes of orthogonal and tilted perfect-fluid models, a $V\to 0$ focusing boundary is a genuine matter singularity: the density diverges and with it a scalar invariant of the Ricci tensor. The explicit Bianchi-I solutions show at the same time that the detailed geometry of the singularity may be highly anisotropic, and that ordinary perfect-fluid matter with $1\leq\gamma<2$ becomes dynamically subdominant to the shear. Stiff matter, $\gamma=2$, is the important exception.

The second lesson is that focusing, geodesic incompleteness and scalar curvature blow-up are not equivalent notions. The Ellis-King Class-B models demonstrate that a homogeneous Cauchy development may encounter a Cauchy horizon while density and scalar curvature invariants remain finite, with the associated curvature behaviour requiring a separate parallelly propagated analysis. Kinematic singularities provide a different possibility: an accelerated tilted fluid congruence becomes incomplete in finite proper time although the spacetime itself remains future geodesically complete. Modern dipole cosmology provides yet another whimper-type regime, with extreme tilt and finite scalar invariants but divergent Riemann components in the metric-comoving tetrad used in the curvature analysis; this frame-dependent blow-up is kept distinct from the parallelly propagated geodesic criterion.

Modern mathematical results have simultaneously shown that the weak or extendible cases are exceptional in important homogeneous families. For generic non-Taub Bianchi IX and non-NUT Bianchi VIII vacuum data, curvature blows up in the incomplete direction, and the corresponding maximal globally hyperbolic developments are $C^2$-inextendible. Ringström’s later analysis of Class-A models placed these results within the dynamical framework of the Bianchi-IX attractor. Thus the historical study of homogeneous cosmological singularities has evolved from a catalogue of exact and qualitative solutions into a subject in which genericity, inextendibility and asymptotic structure can increasingly be formulated as rigorous mathematical theorems.

The present article has deliberately concentrated on what kinds of singular boundaries occur. It has not attempted to describe in detail how the gravitational degrees of freedom evolve as those boundaries are approached. That is the natural subject of *Primer III*. Starting from the ADM and Hamiltonian formulations of homogeneous cosmology, the next article will examine the Kasner map, successive Kasner epochs and eras, the Mixmaster dynamics of Bianchi VIII and IX, modern dynamical-systems formulations, cosmological billiards, and the present status of the BKL conjecture. In this way, the sequence moves from the mathematical definition of singularities (*Primer I*), through their concrete realization in homogeneous cosmology (*Primer II*), to their asymptotic dynamics (*Primer III*).

**Disclaimer/Publisher's Note:** The statements, opinions and data contained in all publications are solely those of the individual author(s) and contributor(s) and not of MDPI and/or the editor(s). MDPI and/or the editor(s) disclaim responsibility for any injury to people or property resulting from any ideas, methods, instructions or products referred to in the content.

**Conflicts of Interest**: The authors declare no conflicts of interest.

**Funding** : This research received no external funding

**Acknowledgments**: The author thanks the anonymous referees for their careful reading of the manuscript and technically valuable remarks, which helped to improved the quality of the work.

**Data** : No new data were created or analyzed in this study. Data sharing is not applicable to this article."